\documentclass[usenatbib]{mn2e}
\usepackage{mathrsfs,amsmath,amssymb,amstext}
\usepackage{lscape}
\usepackage{natbib}
\usepackage{epsfig}
\usepackage{color}
\usepackage{float}
\usepackage{graphicx}
\usepackage{gensymb}
\usepackage[justification=centering]{caption}
\usepackage[flushleft]{threeparttable}
\usepackage[breaklinks,colorlinks,citecolor=blue,linkcolor=magenta]{hyperref} 
\usepackage{footnote}
\makesavenoteenv{tabular}

\usepackage[all]{hypcap} 

\begin{document}

\title[Quantifying  X/Peanut Structures]{Quantifying the (X/Peanut)-Shaped Structure in Edge-on Disc Galaxies: Length, Strength, and Nested Peanuts}
\author[Ciambur \& Graham]{Bogdan C. ~Ciambur,\thanks{E-mail: bciambur@swin.edu.au}  Alister W. ~Graham \\ \\
 Centre for Astrophysics and Supercomputing, Swinburne University of Technology, Hawthorn, VIC 3122, Australia}

\maketitle

\begin{abstract}

X--shaped or peanut--shaped (X/P) bulges are observed in more than 40\% of (nearly) edge-on disc galaxies, though to date a robust method to quantify them is lacking.  Using Fourier harmonics to describe the deviation of galaxy isophotes from ellipses, we demonstrate with a sample of 11 such galaxies (including NGC~128) that the sixth Fourier component ($B_6$) carries physical meaning by tracing this X/P structure.  We introduce five quantitative diagnostics based on the radial $B_6$ profile, namely: its `peak' amplitude (${\it \Pi}_{\rm max}$); the (projected major-axis) `length' where this peak occurs ($R_{{\it \Pi},{\rm max}}$); its vertical `height' above the disc plane ($z_{{\it \Pi},{\rm max}}$); a measure of the $B_6$ profile's integrated `strength' ($S_{\it \Pi}$); and the $B_6$ peak `width' ($W_{\it \Pi}$). We also introduce different `classes' of $B_6$ profile shape. Furthermore, we convincingly detect and measure the properties of multiple (nested) X/P structures in individual galaxies which additionally display the signatures of multiple bars in their surface brightness profiles, thus consolidating further the scenario in which peanuts are associated with bars. We reveal that the peanut parameter space (`length', `strength'  and `height') for real galaxies is not randomly populated, but the three metrics are inter-correlated (both in kpc and disc scale-length $h$). Additionally, the X/P `length' and `strength' appear to correlate with ($v_{\rm rot}/\sigma_{\star}$), lending further support to the notion that peanuts `know' about the galactic disc in which they reside. Such constraints are important for numerical simulations, as they provide a direct link between peanuts and their host disc. Our diagnostics reveal a spectrum of X/P properties and could provide a means of distinguishing between different peanut formation scenarios discussed in the literature. Moreover, nested peanuts, as remnants of bar buckling events, can provide insights into the disc and bar instability history.
\\
\end{abstract}


\newcommand{\elli}{$\textsc{Ellipse}$}
\newcommand{\iso}{$\textsc{Isophote}$}
\newcommand{\bmo}{$\textsc{Bmodel}$}
\newcommand{\cmo}{$\textsc{Cmodel}$}
\newcommand{\ifit}{$\textsc{Isofit}$}
\newcommand{\rf}{\textcolor{red}{reference}}
\newcommand{\rad}{$R_{\rm maj}$}

\section{Introduction}\label{sec:Introduction}

Quite apart from the spheroid-shaped bulges observed at the cores of many disc galaxies, some galaxy bulges display a characteristic bi-lobed, boxy or X-like shape if viewed in projection close to edge-on (for example, the famous IC~4767 -- `the X-galaxy'; \citealt{WhitmoreBell1988}). Shortly after this class of galaxy was noticed (\citealt{BurbidgeBurbidge1959}) they were coined `peanut' bulges by \cite{deVaucouleursdeVaucouleurs1972}, and have since been found to occur in a significant fraction of galaxies (e.g., \citealt{LuettickeDettmarPohlen2000} found that $> 40\%$ of disc galaxies in close to edge-on orientation and ranging from S0 to Sd have (peanut/X)-shaped bulges; see also \citealt{Jarvis1986} and \citealt{ButaEA2015}). Additionally, more recent works suggest that the bulge of our own Galaxy is peanut-shaped ( \citealt{Sellwood1993};  \citealt{DwekEA1995}; \citealt{NessEA2012}; \citealt{WeggEA2015}). 

The early analyses (based on $N$--body simulations) of \cite{CombesSanders1981}, \cite{CombesEA1990} and \cite{RahaEA1991}, suggested that peanuts arise naturally from the thickening and/or buckling of galactic bars, or vertical resonance mechanisms, such as the Inner Lindblad Resonance (ILR). Observational support for the peanut--bar link quickly followed (e.g., \citealt{Shaw1987}; \citealt{DettmarBarteldress1990}; \citealt{KuijkenMerrifield1995}; \citealt{BureauFreeman1999}; \citealt{BureauAthanassoula2005}), the fraction of barred disc galaxies ($\sim$45\%; \citealt{AguerriEA2009}) being found to be close (within a few per cent) to that of X/P hosts. Other than their characteristic morphology, X/P bulges also exhibit specific kinematic signatures in face-on orientation (\citealt{DebattistaEA2005}, \citealt{Mendez-AbreuEA2008}), as well as {\it cylindrical rotation}, i.e., almost constant rotational velocity in the direction perpendicular to the disc plane (e.g., \citealt{MolaeinezhadEA2016} and references therein). The field has advanced significantly, with three--dimensional simulations becoming ever more refined (e.g., \citealt{QuillenEA2014}; \citealt{AthanassoulaEA2015}; \citealt{LiShen2015}) and peanuts being detected at lower inclinations (e.g., \citealt{ErwinDebattista2013}). For a more in-depth overview we refer the reader to the review articles of \cite{LaurikainenSalo2015} and \cite{Athanassoula2016}, and references therein.

Quantifying X/P structures in real galaxies has proved a challenging feat, with most observational studies relying on qualitative methods, such as visual inspection. Several quantitative diagnostics have been attempted, particularly: determining whether isophotes crossing the minor axis are `pinched' or concave (\citealt{WilliamsBureauCappelari2009}), examining one--dimensional cuts parallel to the disc plane (\citealt{DOnofrioEA1999}; \citealt{Athanassoula2005}) or looking for the peanut imprint in the fourth Fourier component ($B_4$) of isophotes (\citealt{BeatonEA2007}; \citealt{ErwinDebattista2013}). However, all of these approaches are problematic. As we will show in Section \ref{sec:analysis}, many peanut galaxies do not exhibit a very prominent X/P feature compared to the disc and spheroid. As such, their photometric structure is dominated by the combination of the latter two, thus most often resulting in straight (parallel to the disc plane) or even $convex$ isophotes which cross the minor axis\footnote{There are of course exceptions, most notably NGC~128, in which the peanut is the dominant component, and the isophotes crossing the minor axis are highly concave.} (where the shape is dominated by the spheroid), despite the presence of an X/P. Taking cuts parallel to the disc plane also works best only for strong peanuts, and further suffers from noise from, e.g., substructure, seeing, etc. Finally, $B_4$ usually measures the boxyness or discyness of isophotes. However, X/P structures are believed to be different to boxy ellipticals (\citealt{CombesEA1990}), and in fact for edge-on galaxies the disc makes the isophotes discy, i.e., when the isophote crosses the $major$ axis, it is highly {\it convex} (not concave, as would be the case for boxy isophotes).

In this work we propose a method which is more accurate, easy to automate, can probe fainter peanuts and is well suited to compare observations (galaxy isophotes) with simulations (projected isodensity contours). \cite{Ciambur2015}, hereafter C15, recently suggested that, just as $B_4$ quantifies boxyness or discyness, so the sixth Fourier component of isophotes ($B_6$) carries physical meaning, capturing the X/P feature. In this paper we show that $B_6$ does trace X/peanut structures remarkably well and provides a wealth of information about them.

The paper is structured as follows. In Section \ref{sec:theory} we define our five diagnostics, based on the radial $B_6$ profile, and we additionally present them in the context of peanut formation theory. Section \ref{sec:data} details our sample of eleven galaxies with known X/P structures, obtained from the literature. Section \ref{sec:analysis} provides a detailed demonstration of X/P structure quantification using a well known peanut galaxy, NGC~128, as well as the results for the full sample. We interpret and discuss our results in Section \ref{sec:discussion}, and finally, we re-iterate our main findings and conclude with Section \ref{sec:conclusions}.
Throughout this paper, we assume a flat Universe cosmology with $\Omega_m$ = 0.27 and H$_0$ = 70 km s$^{-1}$ Mpc$^{-1}$.

\section{Theory and Quantitative X/P Parameters}\label{sec:theory}

Bars in discs are dynamically unstable, and are susceptible to buckling outside of the disc plane. Numerical simulations have shown that bar buckling is one channel through which disc galaxies can develop X/P structures (\citealt{RahaEA1991}; \citealt{MerrittSellwood1994}). In such cases, the bar buckles and forms a peanut in the inner regions (\citealt{AthanassoulaMV2009}) and the stronger the buckling, the more pronounced is the resulting peanut. Another scenario which also arose from $N-$body simulations is that the peanut is actually due to a resonance mechanism (e.g., \citealt{CombesSanders1981}; \citealt{CombesEA1990}), in particular a vertical Inner Lindblad Resonance (ILR). The X/P shape is given by stellar orbits undergoing a 2:1 vertical Lindblad Resonance:

\begin{equation}
\label{equ:resonance}
{\it \Omega_{p,b}} = {\it \Omega} - \frac{\nu_z}{2} , 
\end{equation}

where ${\it \Omega_{p,b}}$ is the pattern rotation frequency of the bar, ${\it \Omega}$ is the circular rotation frequency of stars in the disc, and $\nu_z$ is their vertical oscillation frequency. Essentially, there is a `sweet spot' in radius where the vertical component of stellar orbits is resonant with the bar's pattern speed, and for the 2:1 resonance, the stellar orbits oscillate twice per bar rotation (i.e., each time the two ends of the bar pass underneath). As this is a resonance, this vertical motion enhancement causes the orbits of these stars to `puff' out of the disc plane at these two points, thus leading to the observed peanut shape in an edge-on projection. Finally, \cite{PatsisEA2002b} explore $N$-body simulations where X/P bulges develop without a bar at all.

In order to disentangle the various mechanisms and thus constrain galaxy dynamics, it is obviously interesting to measure a number of peanut properties: the radius where it is most prominent, how high above the disc plane it reaches, and how strong it is, in real galaxies. Our methodology provides a framework for measuring these quantities and thus allows to probe the theory via direct observations, as well as comparing real X/P galaxies among themselves. This framework is based on the azimuthal profiles of galaxy isophotes. Throughout our work, we employed the quasi-elliptical isophote fitting software \ifit, introduced in C15. As detailed in C15, \ifit \ provides a superior description of isophote deviations from pure ellipses (which real galaxy isophotes are observed to have), than its predecessor, \elli. These deviations are mathematically expressed as Fourier harmonics, and in \ifit, they are expressed as a function of the eccentric anomaly ($\psi$) of the ellipse, such that:

\begin{equation}
	R'(\psi) = R(\psi) + \sum_{n} \left[ A_{n} \textrm{sin}(n\psi) + B_{n} \textrm{cos}(n\psi) \right] ,
	\label{equ:FHarmpsi}
\end{equation}

where $R(\psi)$ is the angle-dependent radial co-ordinate of a pure ellipse, $n$ is the harmonic order and $\psi$ is the azimuthal angular co-ordinate (see Figure \ref{fig:peanut_radius}). The angle $\psi$ is related to the usual polar co-ordinate $\phi$ as follows. If a point on a circle is defined in polar co-ordinates ($r,\phi$), and the circle is `flattened' in one direction such that it becomes an ellipse, then the same point has the new co-ordinates ($R(\psi), \psi$). By using the eccentric anomaly as the angular co-ordinate, \ifit\ drastically improves on previous isophote-fitting algorithms, such as \elli. 

Note that generally the harmonic coefficients ($A_n$, $B_n$) can have either units of length (assumed in Equation \ref{equ:FHarmpsi}), or intensity (as \ifit\ uses and outputs), or they can be dimensionless. Fortunately, it is relatively easy to alternate between units if the isophote semi-major axis and local intensity gradient are known. As usually done in the literature, we plot the dimensionless coefficients (i.e., $B_n$ in units of intensity, re-normalised by the isophote semi-major axis ($a$) and local intensity gradient $-\partial I/\partial a$), as given by:

\begin{equation}
B_n \rightarrow- B_n \left( a \frac{\partial I}{\partial a} \right)^{-1}; \;\;\; A_n \rightarrow- A_n \left( a \frac{\partial I}{\partial a} \right)^{-1}.
\label{equ:normalisation}
\end{equation}

\begin{figure}
	\centering
	\includegraphics[width=0.9\columnwidth]{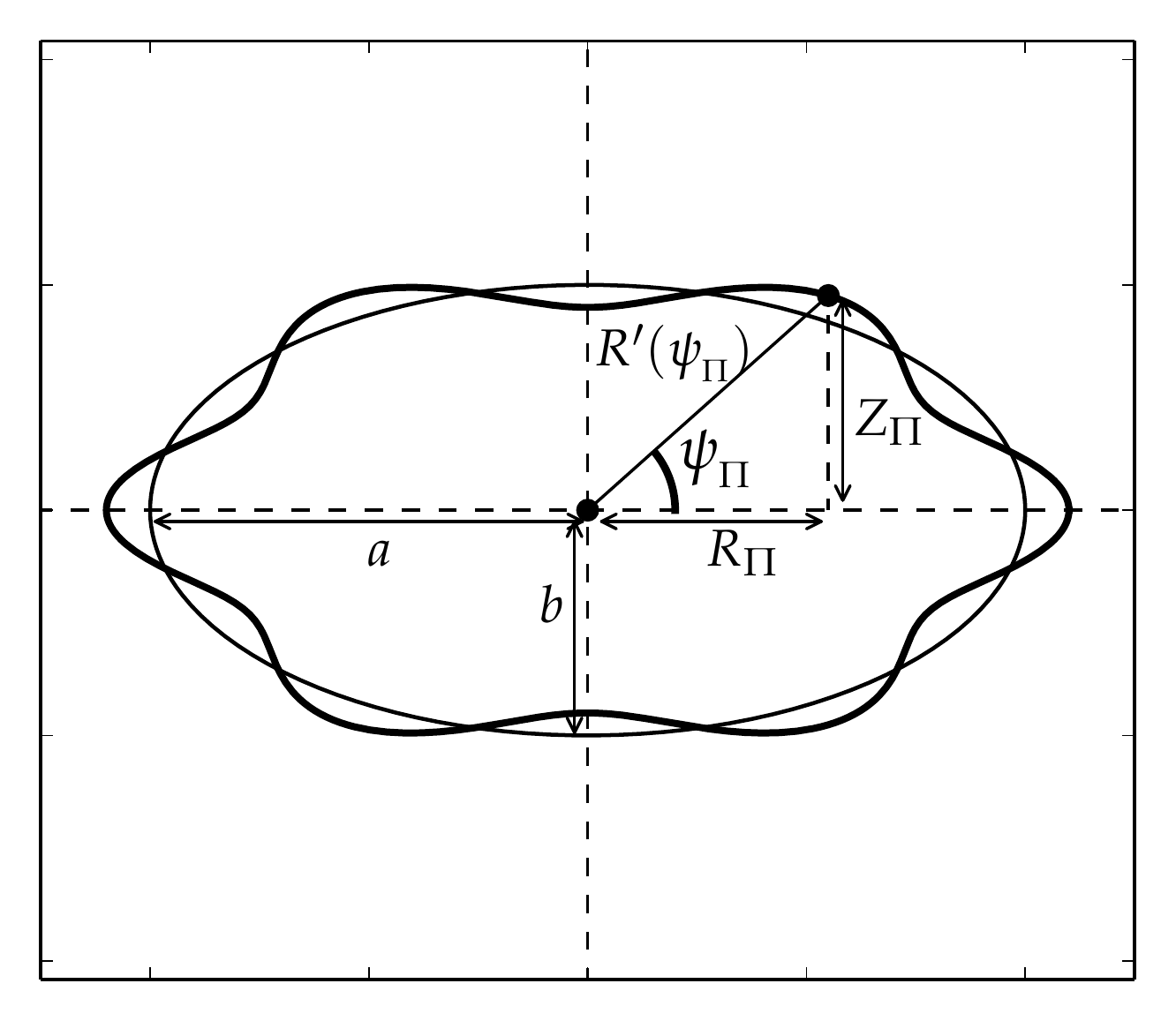}
	\caption{An isophote with the characteristic X/P shape (thick curve), obtained by perturbing an elliptical isophote ($e = 0.4$, thin curve) with the sixth Fourier harmonic ($B_6 = 0.1$). The projected radial `length' of the peanut, $R_{\it \Pi}$, is defined here, as is its projected `height' above the disc plane, $Z_{\it \Pi}$.}
	\label{fig:peanut_radius}
\end{figure}

 In the past, quantitative studies of galactic bars have been performed on similar principles, most notably in \cite{OhtaEA1990}, \citealt{ReganElmegreen1997} and \citealt{AguerriBeckmanPrieto1998} (the latter in particular use diagnostics for bar $length$ and $strength$ derived from isophotal $n=2$ Fourier terms), but see also \cite{Schwarz1984} and \cite{ButaEA2004}. Out of all the Fourier coefficients of Equation \ref{equ:FHarmpsi}, the sixth cosine coefficient $B_6$ is of particular interest in our work quantifying X/peanut-- shaped features. C15 already suggested that $B_6$ may trace the X/P feature in galaxies, and qualitatively explored this possibility (see e.g., their Fig. 4). Here we demonstrate that this is indeed the case, and we define the following peanut diagnostics, derived from the $B_6$ profile (see Figure \ref{fig:b6_expl}):

\begin{enumerate}

\item ${\it \Pi}_{\rm max} (= B_{6,\rm max})$ -- the X/P {\it peak}, equal to the maximum $B_6$ amplitude. The quantity $B_6$, like all the $A_n$ and $B_n$ coefficients, is a function of radius, such that each isophote has its own value.

\item $R_{{\it \Pi},{\rm max}}$ -- the projected X/P {\it length}, equal to the (major-axis) radius where ${\it \Pi}_{\rm max}$ occurs. Note that the true length of a peanut ($l_{{\it \Pi},{\rm max}}$) is only measurable from a galaxy image when the bar is viewed perfectly side-on. As the bar's viewing angle ($\alpha$ in our notation) is unknown, we can only calculate a lower limit of the peanut length, that is, the projected radius $R_{{\it \Pi},{\rm max}} = l_{{\it \Pi},{\rm max}}\,{\rm sin}(\alpha)$ ($\alpha = 0\degree$ for end-on and $90\degree$ for side-on orientation). This latter quantity is straightforward to compute from Equation \ref{equ:FHarmpsi}:

\begin{figure}
	\centering
	\includegraphics[width=1.\columnwidth]{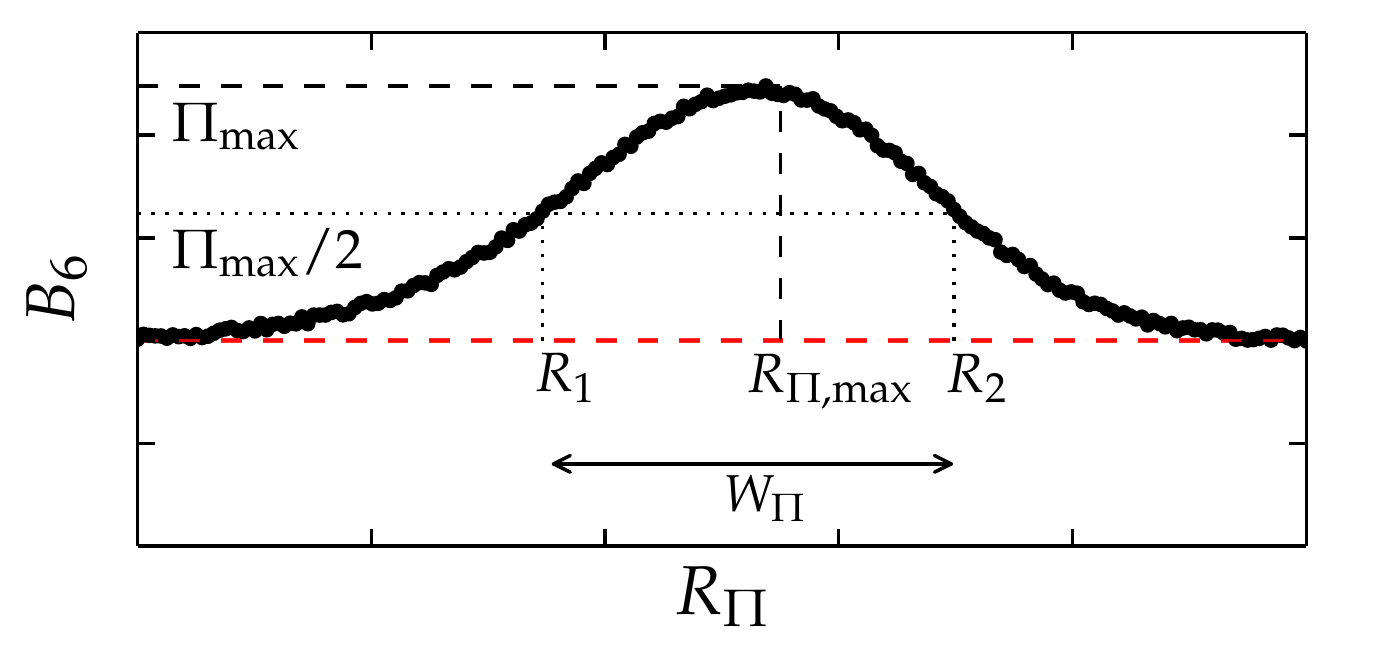}
	\caption{A schematic example of a $B_6(R_{\it \Pi})$ profile (see Figure \ref{fig:peanut_radius} for the definition of $R_{\it \Pi}$), which illustrates the main elements used in defining the five X/P metrics, i.e., the peanut `peak' (${\it \Pi_{\rm max}}$), the projected `length' ($R_{{\it \Pi},{\rm max}}$), the peak `width' ($W_{\it \Pi}$). The peanut `strength' ($S_{\it \Pi}$) is the area under the profile enclosed by $R_1$ and $R_2$. As with the X/P length, the peanut `height' ($z_{{\it \Pi},{\rm max}}$) also corresponds to the isophote with maximum $B_6$ amplitude (${\it \Pi_{\rm max}}$).}
	\label{fig:b6_expl}
\end{figure}

\begin{equation}
\label{equ:r_pi}
R_{{\it \Pi},{\rm max}} = R'(\psi_{_{\it \Pi}}){\rm cos}(\psi_{_{\it \Pi}}) \,,
\end{equation}

calculated at the isophote with maximum $B_6$ amplitude, and where $\psi_{_{\it \Pi}}$ is the eccentric anomaly angle of the peanut, given by:

\begin{equation}
\label{equ:psi}
\psi_{_{\it \Pi}} = -{\rm arctan}\left[  \frac{{\rm tan}(2\pi/6)}{1 - \epsilon} \right] \,,
\end{equation}

with $\epsilon$ being the isophote ellipticity. 

\item $z_{{\it \Pi},{\rm max}}$ -- the X/P {\it height}, equal to the vertical extent above the disc plane computed from the isophote where ${\it \Pi}_{\rm max}$ occurs:

\begin{equation}
\label{equ:zpi}
z_{{\it \Pi},{\rm max}} = \frac{Z_{{\it \Pi},{\rm max}}}{{\rm sin}(i)} = \frac{R'(\psi_{_{\it \Pi}})\,{\rm sin}(\psi_{_{\it \Pi}})}{{\rm sin}(i)} \,,
\end{equation}

where $Z_{{\it \Pi},{\rm max}}$ is the projected height and $i$ is the inclination angle of the galaxy's disc (such that 90$\degree$ corresponds to edge-on). Since X/P structures arise from $vertical$ oscillations ($\perp$ to the disc plane), the observed vertical extent is diminished in proportion to the observed disc inclination. We correct for this effect for non edge-on galaxies in our sample through the use of Equation \ref{equ:zpi}, thus recovering the true vertical extent of the peanut, $z_{{\it \Pi},{\rm max}}$.  We illustrate how $R_{{\it \Pi},{\rm max}}$ and $Z_{{\it \Pi},{\rm max}}$ are derived from an isophote in Figure \ref{fig:peanut_radius}.

\item $S_{\it \Pi}$ -- the integrated {\it strength} of the X/P feature, which we define as:

\begin{equation}
\label{equ:sharpness}
S_{\it \Pi} = 100 \times \int_{R_1}^{R_2} B_6(R) dR \,,
\end{equation}

where $R_1$ and $R_2$ are the (major-axis) radii enclosing the part of the $B_6(R)$ profile higher than the peak half-maximum\footnote{Variations of this can readily be envisaged, and we tested a few but preferred ${\it \Pi}_{\rm max}/2$.} (${\it \Pi}_{\rm max}/2$). \\

\item $W_{\it \Pi}$ -- the X/P {\it peak width}, is the radial extent $R_2 - R_1$.

\end{enumerate}

Although the harmonic profiles are typically plotted as a function of the isophotal semi-major axis, i.e., `$a$' in Figure \ref{fig:peanut_radius}, we plot $B_6$ as a function of $R_{\it \Pi}$ (see Equation \ref{equ:r_pi} and Figure \ref{fig:peanut_radius}).

In our analysis, an X/P feature can be `strong' in two ways: when it displays a high $B_6$ amplitude, or if the peak in the $B_6$ profile is extended over a large radial range (i.e., a large $W_{\it \Pi}$). This will be more clear when we present actual $B_6$ profiles in the following sections, and is illustrated here schematically in Figure \ref{fig:b6_expl}. Both of these cases are accounted for in our definition of the peanut \textit{strength} diagnostic, $S_{\it{\Pi}}$. We chose to perform the integral in Equation \ref{equ:sharpness} within the half-maximum limits so that the X/P feature is well separated from other potential components or the noise in the $B_6$ profile, as $B_6$ approaches zero. Throughout this work, we provide the X/P length, height, strength and peak width both in the usual angular units (arcsec), in kiloparsecs and in units of the disc exponential scale length $h$. We prefer the latter, since it is more insightful from a galaxy structure perspective, and is independent of the distance to the galaxy.

Perhaps the most important advantage of our method is that it is simple, practical, and constitutes an $objective$ and $accurate$ way of measuring X/P structures. Moreover, it is not restricted to the strongest and most visibly obvious X/P bulges, but can detect and measure weaker structures, in regions of the galaxy where additional components (such as the bulge or the disc) dominate the isophotes. 

\section{Data}\label{sec:data}

It is difficult to decide upon the optimal observational data for edge-on peanut galaxies. An important consideration is dust obscuration, which is likely to be a problem for peanut galaxies because they are disc galaxies viewed, in the majority of cases, at high inclination or close to edge-on. Therefore, were their discs to have embedded dust lanes (which is quite common in actively star-forming spirals), the undesired obscuration would have a maximal impact. Opting for near-infrared (NIR) imaging has the advantage of revealing X/P structures which may be contaminated (or even completely obscured) in the visible bands by dust (\citealt{QuillenEA1997}). However, the disadvantage of NIR data is a broader point spread function (PSF), often with strong Airy rings. Should a galaxy contain a large number of point-like or (small) resolved objects (such as a bright AGN, star clusters, etc.), their light is spread by the instrumental PSF into the diffuse galaxy light (i.e., the peanut, bulge, disc) and thus introduces noise/bias in the true shape of these components. To diminish this effect, shorter wavelengths are preferable, particularly since determining the shape of the peanut accurately makes it possible to detect weak or nested peanuts, which would otherwise be lost in the noise.

\begin{table*}
\centering
\begin{minipage}{35mm}
\caption{The X/P Sample}
\centerline{
	\begin{tabular}{l l l c c c c c}
	\hline 
	Galaxy (peanut) & Telescope (filter) & Type & Scale& Inclination $i$	 & $v_{\rm rot}$ & $\sigma_{\star}$ & $M_K$\\
	 &  &  &  [pc arcsec$^{-1}$]& [$\degree$]	 & [km\,s$^{-1}$] & [km\,s$^{-1}$] & [mag]\\
	\hline \hline \\
	NGC~128 $(a,b)$ & $HST$, $NIC3$ ($F160W$) & S0 pec  & 288 & 78 & 180$\pm$40$^{*}$ & 211$\pm$32 & -25.35$\pm$0.15 \\
	NGC~128 $(b)$ & SDSS ($r$)\\
	NGC~678 & $Spitzer$ ($3.6\mu$) & SB(s)b?  & 193 &  90 & 195$\pm$29 & 133$\pm$20 & -24.29$\pm$0.15 \\
	NGC~2549 ($a$) & $HST$ ($F702W$)  &SA0$^{0}$(r) & \;\;95 & 75 & 160$\pm$10$^{*}$ & 143$\pm$22 & -23.40$\pm$0.15\\
	NGC~2549 ($b$) & SDSS ($r$)  \\
	NGC~2654 & $Spitzer$ ($3.6\mu$) & SBab:  & 117 & 81 & 236$\pm$10 & --- & -23.45$\pm$0.15 \\
	NGC~2683 & $Spitzer$ ($3.6\mu$) & SA(rs)b & \;\;32 & 82 & 218$\pm$7\;\;  & 118$\pm$18 & -22.75$\pm$0.15 \\ 
	NGC~3628  & $Spitzer$ ($4.5\mu$) & Sb pec  &  \;\;43 & 90 & 230$\pm$5\;\; & \;\;81$\pm$12 & -23.67$\pm$0.15 \\
	NGC~4111  & SDSS ($i$) &SA0$^{+}$(r):  & \;\;75 & 90 & \;\;89$\pm$6\;\; & 147$\pm$22 & -23.41$\pm$0.15\\
	NGC~4469  & SDSS ($z$)& SB(s)0/a?   & \;\;19 & 75 & \;\;18$\pm$9\;\; & 107$\pm$16 & -19.93$\pm$0.20\\
	NGC~4710  & $Spitzer$ ($3.6\mu$) & SA0$^{+}$(r)?  & \;\;71 & 90 & 150$\pm$36 & 117$\pm$18 & -23.25$\pm$0.15 \\
	NGC~7332 & SDSS ($r$) & S0 pec  & \;\;96 & 80 & 186$\pm$28 & 128$\pm$19 & -23.47$\pm$0.15 \\
	ESO~443-042 & $Spitzer$ ($4.5\mu$) & Sb:  & 175 & 90 & 191$\pm$8\;\; & --- & -23.11$\pm$0.15 \\
	\hline
	$^{*} v_{\rm rot}$ of stars
	\end{tabular}
	\label{tab:xpsampleinfo}
}

\end{minipage}
\end{table*}

We compiled a sample of 11 nearly edge-on galaxies with known X/P structures from the literature (\citealt{Sandage1961}; \citealt{Tsikoudi1980}; \citealt{Jarvis1986}; \citealt{deSouzadosAnjos1987}; \citealt{Shaw1987}; \citealt{ShawDettmarBarteldress1990}). We sourced the imaging for our sample from three archives, namely $i)$ the {\it Hubble Legacy Archive\footnote{http://hla.stsci.edu}} for visible and NIR imaging taken with the {\it Hubble Space Telescope} ($HST$) $ii)$ Data Release 9 (DR9, \citealt{AhnEA2012}) of the Sloan Digital Sky Survey\footnote{https://www.sdss3.org/dr9/} (SDSS), where we generally opted for the redder bands ($r, i, z$) and $iii)$ the {\it $Spitzer$ Survey of Stellar Structure in Galaxies\footnote{http://irsa.ipac.caltech.edu/data/SPITZER/S4G/}} (S$^{4}$G, \citealt{ShethEA2010}, \citealt{ButaEA2015}) for NIR imaging (3.6$\mu$ and 4.5$\mu$). We decided on the optimal source on a case-by-case basis, depending on the image availability, level of dust obscuration and field of view size. For NGC~128 and NGC~2549, we made use of more than one image. 

Table \ref{tab:xpsampleinfo} lists the full sample, together with the source archive and bandpass (filter), as well as the morphological classification from the {\it Third Reference Catalogue of Bright Galaxies}\footnote{http://vizier.u-strasbg.fr} \citep{deVaucouleursEA1991}, hereafter RC3. The inclination angle $i$ of each galaxy is also listed, and was calculated from the axis ratio reported in \cite{SaloEA2015} for the S$^4$G galaxies, and in RC3 for the remainder, while assuming an intrinsic thickness of an edge-on disc of 0.22 (\citealt{UnterbornRyden2008}). The intrinsic thickness is degenerate with the inclination: when a disc approaches an edge-on viewing angle, its axis ratio is dominated by its natural thickness. For this reason, all our galaxies with an RC3 value of $b/a < 0.22$ are assumed to be exactly edge-on ($i = 90\degree$).  Additionally, Table \ref{tab:xpsampleinfo} lists the physical scale (pc arcsec$^{-1}$) of each image, corrected for infall into the Virgo Cluster and the Great Attractor, as provided by the NASA/IPAC {\it Extragalactic Database\footnote{https://ned.ipac.caltech.edu}} (NED), plus the $2MASS$\footnote{http://www.ipac.caltech.edu/2mass/} $K$--band magnitude ($M_K$) of each galaxy, as listed in NED. Finally the gas rotational velocity ($v_{\rm rot}$) and the stellar velocity dispersion ($\sigma_{\star}$) were retrieved from the $HyperLeda$\footnote{http://leda.univ-lyon1.fr} (\citealt{MakarovEA2014}) database and are also listed in Table \ref{tab:xpsampleinfo} for each galaxy. We used the $v_{\rm rot}$ of the gas reported by \cite{GiovanelliEA1986}, \cite{GiovanelliEA2007}, \cite{HaynesEA2011} and \cite{BraunEA2003} for NGC~678, NGC~4710, NGC~4469 and NGC 7332, respectively; the $stellar$ rotational velocity reported by \cite{BertolaCapaccioli1977} and \cite{SimienPrugniel1997} for NGC~128 and NGC2549; and gas rotational velocities values reported by \cite{CourtoisEA2009} for the rest of the sample.

In Table \ref{tab:xpsampleinfo}, those galaxies where an inner, nested peanut was found, are henceforth labelled with $(a)$ for the inner peanut and $(b)$ for the outer.

\section{Analysis}\label{sec:analysis}

\subsection{Methodology: Modelling a Strong Peanut -- NGC~128}\label{sec:analysis - methodology}

Here we demonstrate our analysis of X/P bulges on the interesting, well known peanut galaxy NGC~128. This peculiar lenticular galaxy displays a very pronounced X/P structure (Figure \ref{fig:n128_SDSS}) and is in fact the first galaxy where such a feature was noted (\citealt{BurbidgeBurbidge1959}).

\begin{figure}
	\centering
	\includegraphics[width=1.\columnwidth]{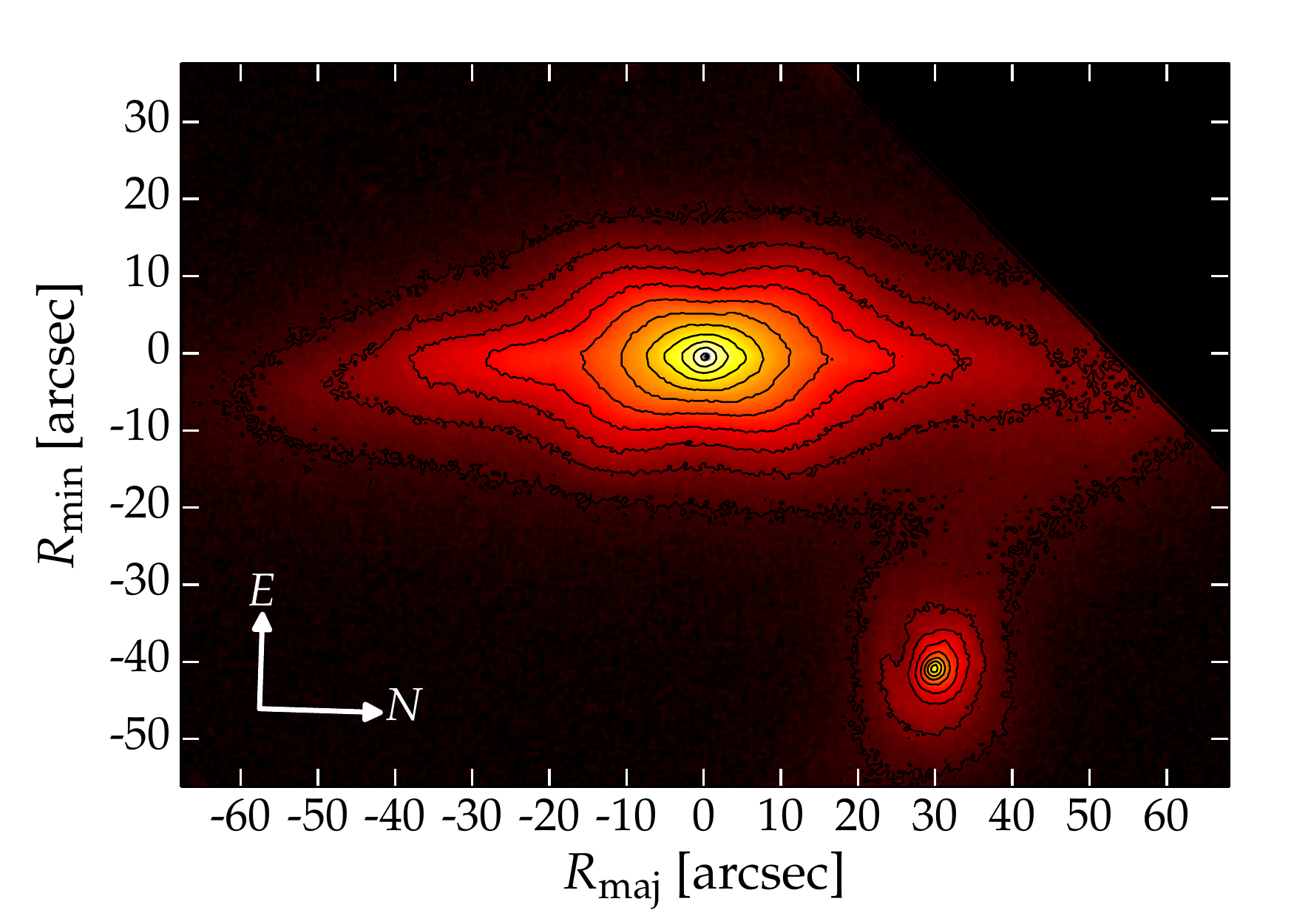}
	\caption{NGC~128 in the SDSS $r$-band. This galaxy has a peculiar disc morphology, particularly at $R \gtrsim 40 \arcsec$, which is likely related to its interaction with the companion galaxy NGC~127, also displayed here (to the West).}
	\label{fig:n128_SDSS}
\end{figure}

The first image that we modelled was the SDSS $r-$band image. Using \ifit, nested isophotes were fit at increasing major-axis steps, each having full freedom in varying their geometric parameters, such as centre position, ellipticity ($\epsilon$) and position angle ($PA$). \ifit\ distorts the purely elliptical shapes at each radial location through Fourier terms (Equation \ref{equ:FHarmpsi}) whose number (1 to $n$) are chosen at the start of the run. The modes chosen for our analysis were $n=2,3,4,6,8,$ and $10$. Briefly, $n=2,3,$ and $4$ are the standard harmonics used in the literature ($n=$2 fits the $\epsilon$ and $PA$, $n=$3 fits for asymmetry and $n=$4 for boxyness/discyness), whereas $n=6,8,10$ are higher order terms which in general are negligible but for peanut galaxies are relevant ($n=6$ is in fact crucial since this is the harmonic which describes the X/P feature, as we will show). Because odd-numbered harmonics capture ever more complicated asymmetries, we did not include $n > 3$ odd terms. 

\begin{figure}
	\centering
	\includegraphics[width=1.\columnwidth]{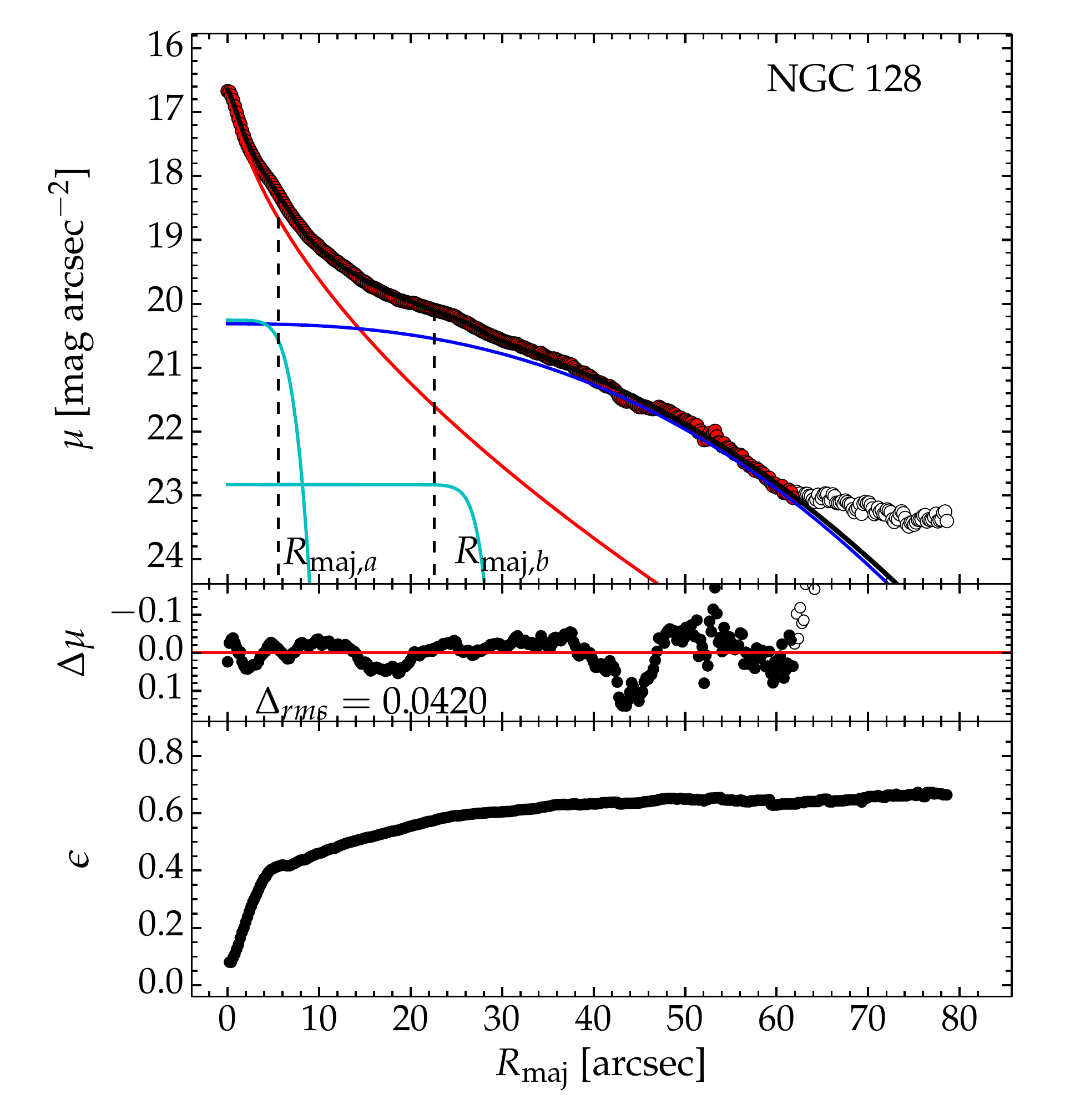}
	\caption{\textit{Top:} The surface brightness profile of NGC~128, extracted from the SDSS $r$--band image, and decomposed into its constituent components: spheroid (red), edge-on disc (blue) and two bars (cyan). The dashed vertical lines correspond to the semi-major axis lengths of the isophotes associated with the inner ($a$) and outer ($b$) X/P structures. \textit{Middle:} The residual profile (data -- model). \textit{Bottom:} The isophote ellipticity profile.}
	\label{fig:n128_decomp}
\end{figure}

The first aspect analysed from our isophotal study was the major axis surface brightness profile $\mu(R_{\rm maj})$, which we subsequently decomposed into galaxy components. There is a long history of using analytical functions to describe the radial surface brightness profiles of galaxies and their components (see \citealt{Graham2013} for a review), with multi-component ($\ge 2$) decompositions routinely fit these days (e.g., \citealt{LaurikainenEA2014}; \citealt{SavorgnanGraham2016}). The decomposition was performed with the software {\sc profiler} (\citealt{CiamburIP}). {\sc profiler} constructs a model $\mu(R_{\rm maj})$ from user-defined analytical functions (such as S\'ersic, exponential, Gaussian, etc.) each of which is intended to describe a particular photometric component (e.g., the disc, the bulge). The model is built by adding together all the components and then convolving the resulting profile with the instrumental point spread function (PSF). This process is iterated until the best-fitting solution is found. {\sc profiler} employs the Levenberg-Marquardt \citep{Marquardt1963} minimisation algorithm, as well as a hybrid (fast Fourier transform -based + direct) convolution scheme. {\sc Profiler} has many desirable features, such as a fast minimisation process (chiefly due to its efficient convolution scheme), an intuitive graphical user interface, and several options for the choice of PSF profile, specifically Gaussian or Moffat (\citealt{Moffat1969}) functions, or any user-provided profile in the form of a table of values. For NGC~128, the PSF was characterised by fitting Gaussian profiles to bright unsaturated stars in the image, with the IRAF task {\sc Imexamine}.

\begin{figure}
	\centering
	\includegraphics[width=1.\columnwidth]{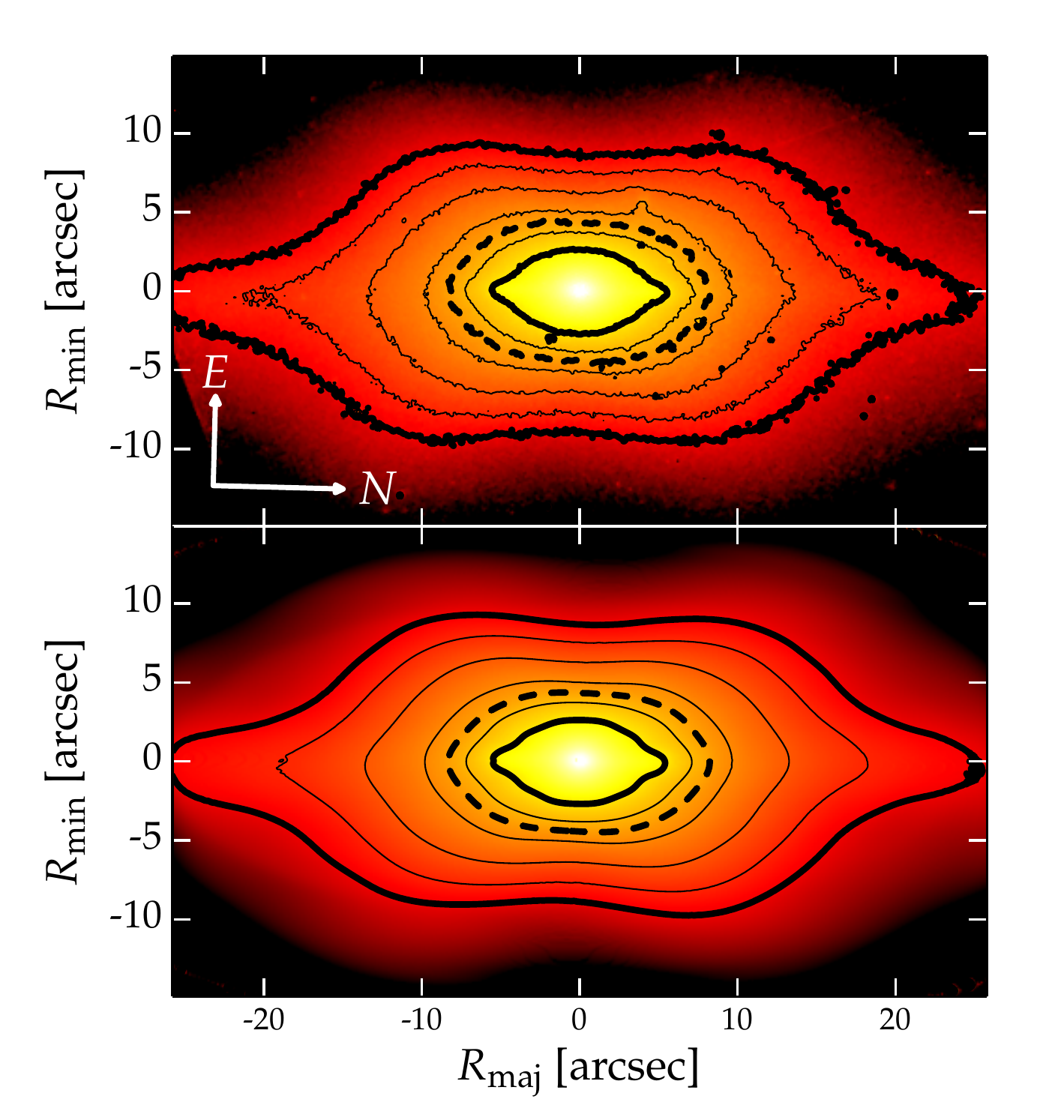}
	\caption{The $HST/NIC 3$ image of NGC~128 (top) and the reconstruction with {\sc Isofit/Cmodel} (bottom). The residual image resulting from subtracting the reconstruction from the image can be seen in the lower panel of Figure \ref{fig:n128_residuals}. Both panels have contours overlayed (at the same intensity levels), to highlight the isophote shapes. The two solid thick contours represent the isophotes corresponding to the two peaks in the $B_6$ harmonic amplitude (see Figure \ref{fig:n128_coeff_HST}), while the thick dashed contours represent the isophote with the minimum $B_6$ amplitude between the two peaks.}
	\label{fig:n128_img}
\end{figure}

The decomposition of NGC~128's surface brightness profile is shown in the top panel of Figure \ref{fig:n128_decomp}. In addition to the spheroid component (red curve) and inclined disc component (blue curve), the profile also displays the signatures of two bar components, plotted in cyan. While multiple, nested, bars are known to occur in some galaxies (\citealt{FriedliEA1996}, \citealt{ErwinEA2001}, \citealt{ErwinEA2011}), this instance is particularly interesting as the bulge of NGC~128 is peanut-shaped, and X/P structures are known to be associated with bars. \cite{Martinez-ValpuestaShlosmanHeller2006} show how, in their simulations, bars can undergo recurrent buckling events, each giving rise to a progressively longer peanut-shaped structure, but the X/P structures themselves do not co-exist (they occur as single structures at different times, not a nested structure). For NGC~128 though, as we will show below, each of the two co-existing bars has its own corresponding peanut. 

The second image that we investigated was a higher-resolution image of NGC~128, taken with the $NICMOS-3$ ($NIC$3) instrument on board the $HST$. While the field-of-view is small, the better spatial resolution at smaller radii is ideal to showcase the X/P feature of this galaxy, and measure it more accurately than from the SDSS $r$--band. Again, the modelling was performed with \ifit, and additionally, a two-dimensional reconstruction of the image was generated with the IRAF task \cmo\; (also introduced in C15). NGC~128 and the reconstructed image are displayed in Figure \ref{fig:n128_img}. The residual image (data minus reconstruction) is shown in Figure \ref{fig:n128_residuals}.

\begin{figure*}
	\centering
	\includegraphics[width=0.7\textwidth]{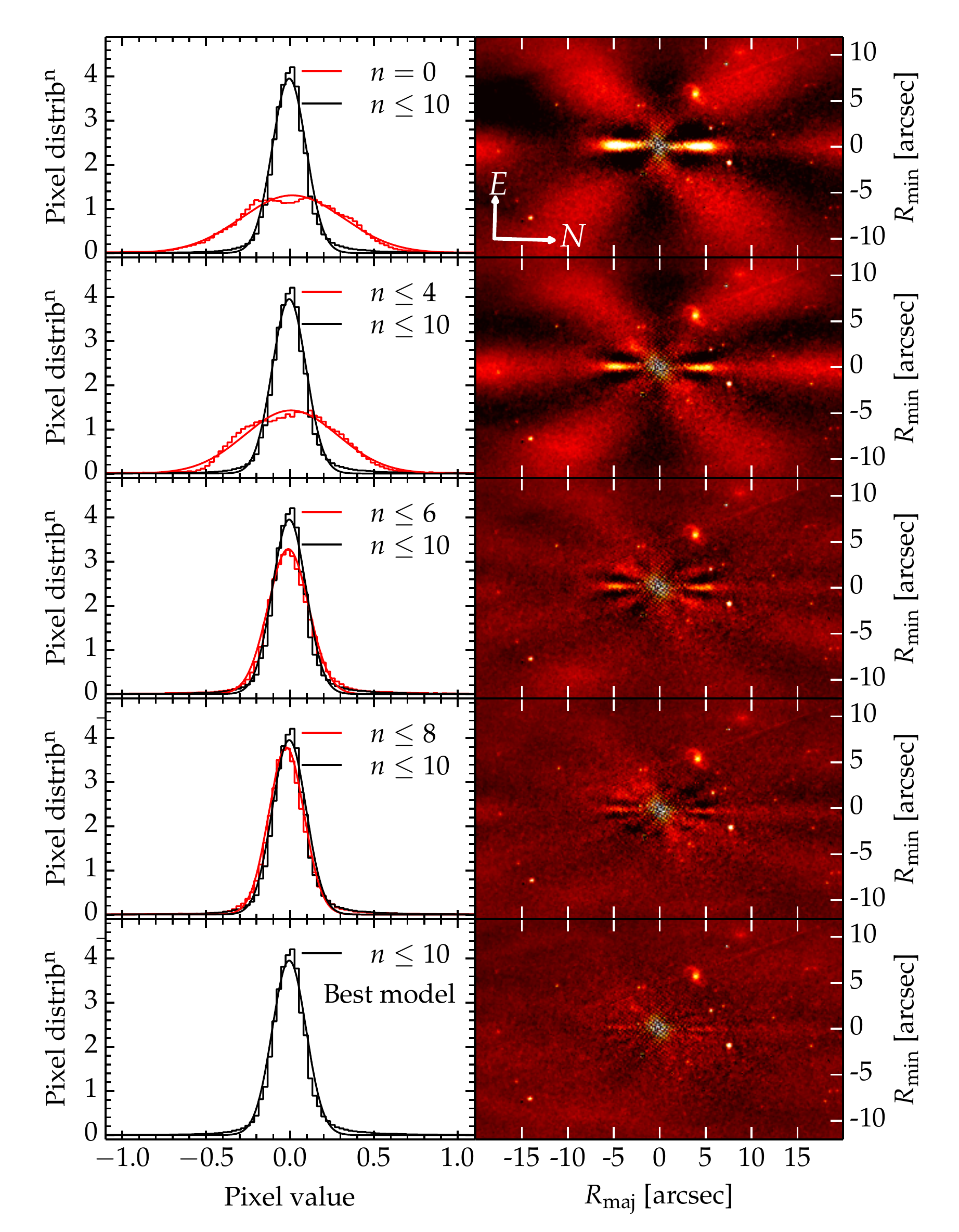}
	\caption{Finding the harmonic order which captures the X/P feature in NGC~128, based on the $NIC 3$ image. \textit{Left panels} -- the red curves represent the pixel distribution in the residual image for five upper limits in the number of harmonic terms, from top to bottom: $n = 0$, i.e., no harmonic terms, purely elliptical isophotes (top), $n \leq 4$ (middle), $n \leq 6$, $n \leq 8$ and our preferred model, $n \leq 10$ (bottom). The black curve is the same in all five panels, and represents the pixel distribution corresponding to the residual image from the highest-quality model (i.e., the bottom panel), which corresponds to the reconstruction in Figure \ref{fig:n128_img} and is adopted troughout our work. \textit{Right panels} -- residual profiles associated with the five cases considered on the left-hand side.}
	\label{fig:n128_residuals}
\end{figure*}

If we were to generate a perfect model of the galaxy, then subtracting it from the image would produce a residual map in which the distribution of pixel values is centred on zero, and has some inevitable dispersion due to the noise in the data. However, failing to adequately capture a given photometric component leads to systematics in the residual, which broaden the pixel distribution and cause it to depart from a (narrow) Gaussian shape. We can discern which is the harmonic that captures the X-shape/peanut-shape from Figure \ref{fig:n128_residuals}. The figure shows the distribution of pixel values in the residual image (left-hand panels) and the residual image itself (right-hand panels), for five values of the maximum harmonic order chosen in \ifit. The first (top) case corresponds to purely elliptical isophotes ($n = 0$)\footnote{Strictly speaking, \ifit\; does use the $n=2$ harmonic to $adjust$ the ellipticity ($\epsilon$) and position angle ($PA$) of ellipses as they are fitted. However, once adjusted, the old ellipses distorted by $n=2$ become new ellipses with no distortions ($n=0$) but updated $\epsilon$ and $PA$.}, for which the residual image displays obvious (and essentially expected) X-shaped systematics. The latter are reflected in the relatively broad dispersion of the pixel distribution (the red curve), which also appears to be bi-modal: the bright features are compensated by dark features, each giving rise to its own peak. It is notable that upgrading the model's level of sophistication to the use of $n \leq 4$, i.e., the standard boxyness/discyness terms, does little to nothing to improve the fit. The dispersion is just as broad, still bi-modal, and the X-shape just as pronounced (Figure \ref{fig:n128_residuals}, middle panels). This clearly shows that, for edge-on galaxies, the X/P feature $cannot$ be quantified with the $B_4$ harmonic term. It is only when the sixth harmonic is included that  the pixel distribution suddenly approaches a Gaussian shape, with a dispersion 2.3 times narrower than the $n \leq 4$ case (2.5 times narrower than for the pure ellipse model), and the X-shaped feature is adequately captured. The pixel distribution resulting from our highest-quality model (corresponding to $n \leq 10$) is overplotted in black in all panels of Figure \ref{fig:n128_residuals}. In passing, we note how \ifit/\cmo\ enable the easy visual identificationof a background spiral galaxy in the image of NGC~128.

The role played by $B_6$ is even more apparent when we plot the radial profiles of the harmonic coefficients. We show the even-$n$ cosine coefficients output by \ifit\; in Figure \ref{fig:n128_coeff_HST}, for the range of harmonic orders $4 \leqslant n \leqslant 10$. 

\begin{figure}
	\centering
	\includegraphics[width=1.\columnwidth]{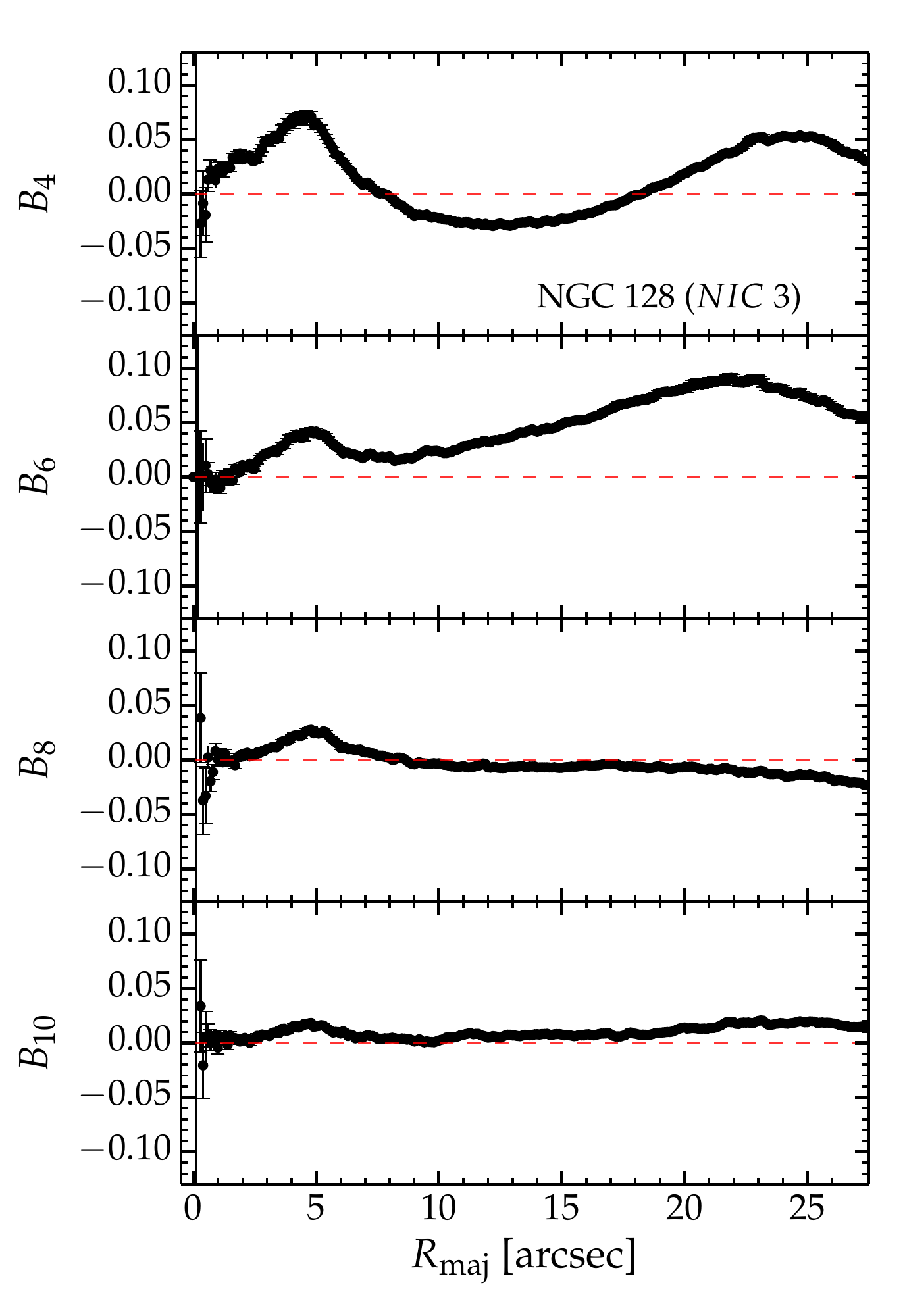}
	\caption{Dimensionless (see Equation \ref{equ:normalisation}) even harmonic cosine coefficients ($B_n$) of order $4 \leq n \leq 10$, plotted as a function of isophote semi-major axis. $B_4$ represents boxyness/discyness while $B_6$ captures the peanut shape and dominates in amplitude over the other harmonic orders at $R_{\rm maj} \sim 10 - 25$ arcsec (see also Figure \ref{fig:n128_coeff_SDSS}), where the X/P feature is most prominent. The higher orders have comparatively low amplitudes, and serve to refine the final isophote shape through small peturbations to the elliptical shape. Data extracted from the $NIC 3$ image (Figure \ref{fig:n128_img}).}
	\label{fig:n128_coeff_HST}
\end{figure}

In Figure \ref{fig:n128_coeff_HST}, $B_4$ (which models boxyness/discyness, as usual) contributes significanly to the isophote shape. The $B_4$ profile indicates a transition from discy ($B_4 > 0$) to boxy ($B_4 < 0$) to discy again with increasing $R_{\rm maj}$. However, it is not this harmonic which captures the X/P feature, but in fact the $B_6$ harmonic (Figure \ref{fig:n128_residuals}). It is clear that $B_6$ dominates over all other coefficients at $R_{\rm maj} \gtrsim 20\arcsec$. This is in fact the locus where the peanut is most prominent, as we show in Figure \ref{fig:n128_img} through the outer thick contour (which corresponds to ${\it \Pi}_{\rm max}$) overlayed on the data and the reconstructed image. As $R_{\rm maj}$ decreases, so does the $B_6$ amplitude, mirroring the isophotes becoming more elliptical (slightly boxy) and the peanut shape less and less apparent. The higher-order harmonics are comparatively low-level, and most likely only serve to refine the final isophote shape, as we shall discuss below.

\begin{figure}
	\centering
	\includegraphics[width=1.05\columnwidth]{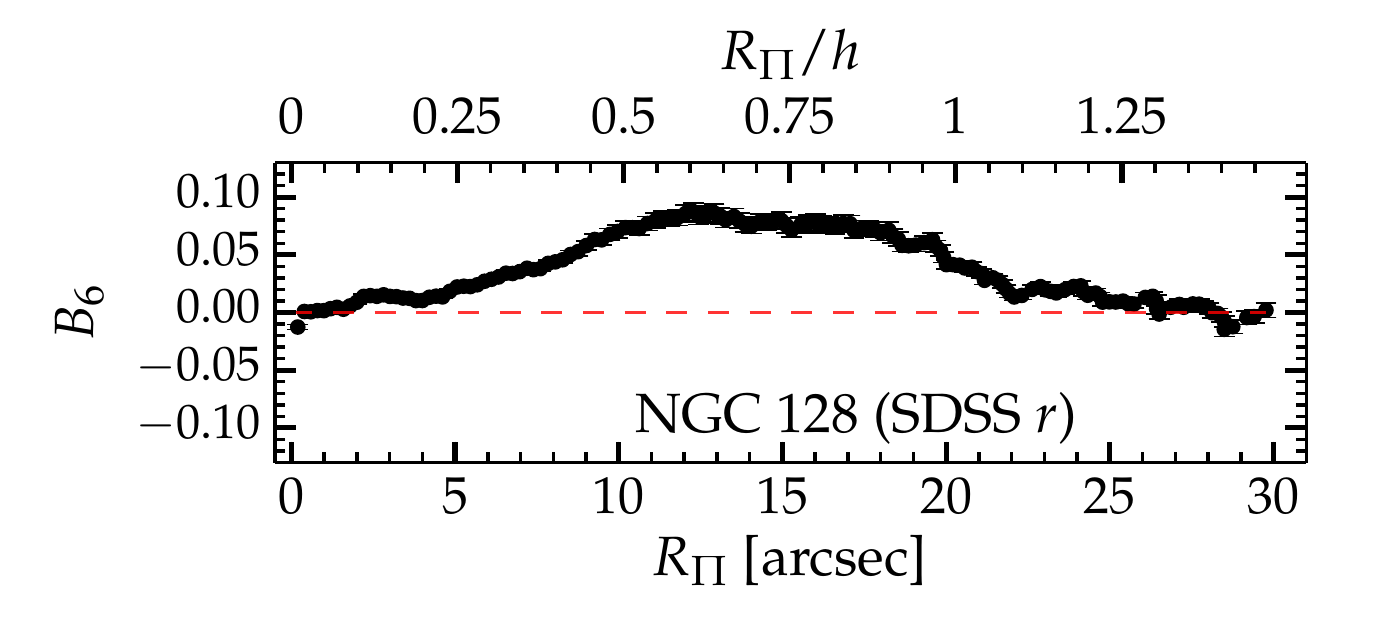}
	\caption{Extended $B_6$ profile of NGC~128, derived from the SDSS $r -$band image (Figure \ref{fig:n128_SDSS}).}
	\label{fig:n128_coeff_SDSS}
\end{figure}

It is worth noting that at \rad\, $\sim 5\arcsec$ there is another peak in the $B_6$ profile. While this is a radius where the isophote has contributions from other harmonic orders, it is likely that there is in fact a second, nested X/P structure in these inner regions. The use of \ifit, therefore, is essential in identifying and measuring very faint X/P structures that would otherwise not be detected. 

Curiously, a faint 8-prong pattern is evident in the central $\sim$ 6$\arcsec$ of the residual image, when employing the $n\leq 6$ model (Figure \ref{fig:n128_residuals}, third panel down). As the $n=6$ harmonic describes the X/P shape, the $n=8$ term then captures this fainter 8-spoked feature (see Figure \ref{fig:n128_coeff_HST}, third panel down). One can speculate that the latter pattern may reflect that in the central $\sim 6\arcsec$, the spheroid component contributes to the isophote shapes along the $minor$-axis, and thus the spheroid, disc an peanut together induce a total of eight `bumps' in the shapes of the inner isophotes: two from the disc (along the major-axis), two from the bulge (along the minor-axis) and four from the peanut's X-like `arms'. This pattern might also be caused by the fact that the two nested X shapes are slightly offset from each other, i.e. they do not necessarily have the same point of origin and certainly do not have the same opening angle $\psi$ of the X `arm' relative to the major-axis (see Table \ref{tab:sample}). This offset may therefore cause these low-level, leftover systematics, which are corrected by the higher order harmonics.

The five X/P diagnostics of NGC~128 were computed from both the $HST$- and SDSS- derived $B_6$ profiles (Figures \ref{fig:n128_coeff_HST} and \ref{fig:n128_coeff_SDSS}). This was necessary because, due to the small field of view of the $HST$ image, the integrated strength of the main (outer) peanut could not be measured reliably ($R_2$ from Equation \ref{equ:sharpness} corresponds to an isophote semi-major axis of $\sim 38\arcsec$, outside of the $NIC$3 CCD). It also provided a consistency check for ${\it \Pi}_{\rm max}$,  $R_{{\it \Pi},{\rm max}}$ and $z_{{\it \Pi},{\rm max}}$ from two different images of the same galaxy. In addition to the main outer peanut, we also measured the five quantities for the inner peanut. The results are tabulated in Tables \ref{tab:sample} and \ref{tab:sample_arcsec-h}.

Figurs \ref{fig:n128_decomp} and \ref{fig:n128_coeff_HST} nicely show how the two X/P structures of NGC~128 are associated with its two bars. This is indicated through the major-axis radii of the isophotes corresponding to the two peanuts of this galaxy, namely $R_{{\rm maj},a}$ for the inner peanut and $R_{{\rm maj},b}$ for the outer. We draw the reader's attention to the fact that, for a given isophote, the projected peanut radius $R_{\it \Pi}$ is different to the isophote's major-axis radius $R_{\rm maj}$, the latter being always longer than the former (see Figure \ref{fig:peanut_radius}). In Figures \ref{fig:n128_decomp} and \ref{fig:n2549_decomp} we follow the standard practice of plotting the surface brightness profile as a function of isophote semi-major axis, i.e., $R_{\rm maj}$. For consistency, we indicate in these two plots the $R_{\rm maj}$ of the isophotes having the maximum $B_6$ value, corresponding to the two peanuts. These two radii clearly correspond visually to approximately the ends of the inner and outer bars (i.e., the radii where the bar profiles cease to be flat and start to drop off in surface brightness), respectively, though we recognise that the actual peanut $lengths$ are shorter than the bars, which is in agreement with simulations (see the review article by \citealt{Athanassoula2016}, and also \citealt{AthanassoulaMisiriotis2002}, \citealt{Martinez-ValpuestaShlosmanHeller2006}).

\subsection{The Full Galaxy Sample}\label{sec:analysis-sample}

Having now detailed the analysis and quantification of X/P structures in our case-study of NGC~128, we now present the remaining galaxies. While NGC~128 hosts a remarkably strong and obvious peanut structure, this is not necessarily the case for all such galaxies (moreover, its inner peanut was hidden under the luminosity of the bulge and disc). Peanut bulges come in a wide range of shapes and sizes: some puffing up strongly outside the disc plane and showing a clear X-shape (e.g., NGC~3628, ESO~443-042), others (e.g., NGC~2654, NGC~2683) displaying so-called $spurs$ (\citealt{ErwinDebattista2013}), some barely perturbing the discy/lenticular shape of their host (e.g. NGC~2549 ($b$)) while others showing peanut-like isophotes `pinched' along the minor axis (e.g., NGC~128, NGC~4469). Nevertheless, our method of isophotal analysis proved robust for all of these cases, as the peanut invariably left its imprint on the radial $B_6$ profiles of these galaxies.

The results of the isophotal analysis (image, reconstruction, residual and $B_6$ profile) for each individual galaxy are shown in Appendix A (Figure \ref{fig:sample-1}), while we report the peanut diagnostics in Table \ref{tab:sample}. Before proceeding, provide some additional remarks concerning some of the galaxies in our sample. 

\begin{figure}
	\centering
	\includegraphics[width=1.\columnwidth]{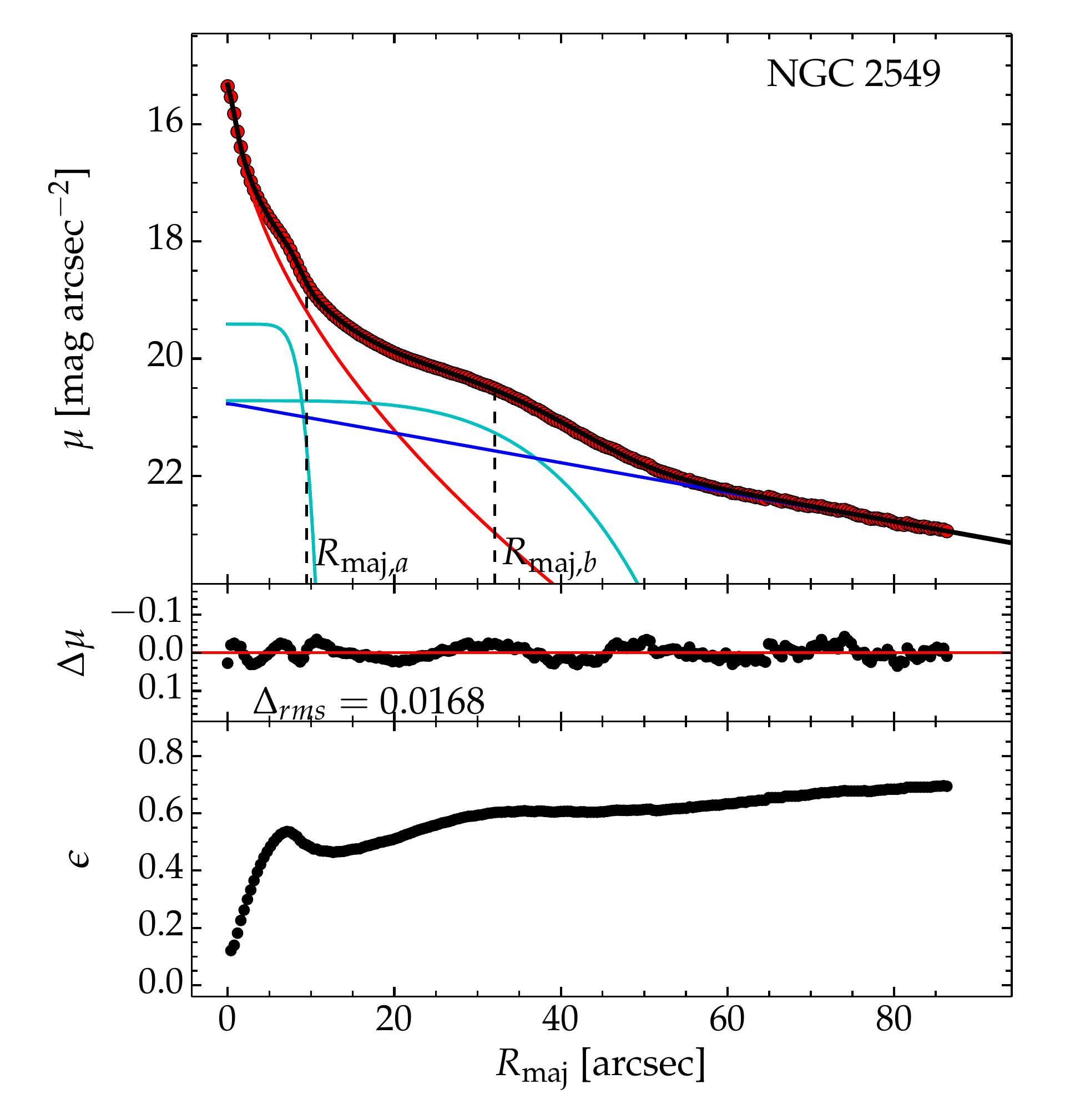}
	\caption{The decomposition of the SDSS $r -$band surface brightness profile of NGC~2549. Apart from a S\'ersic (spheroid) and an exponential (disc) component, the profile also shows the signatures of two bars, modelled with S\'ersic functions of low $n$ ($\sim 0.1 - 0.2$), and plotted as the cyan curves. As in Figure \ref{fig:n128_decomp}, the major-axis radii of the isophotes associated with the two peanuts ($R_{{\rm maj},a}$ and $R_{{\rm maj},b}$) are indicated as vertical dashed lines.}
	\label{fig:n2549_decomp}
\end{figure}

\begin{table*}
\begin{minipage}{50mm}
\caption{The X/P Diagnostics, in kpc}
\centerline{
	\begin{tabular}{l c c c c c c c l}
	\hline 
	Galaxy  & $h$  & ${\it \Pi}_{\rm max}$ & $R_{{\it \Pi},{\rm max}}$ & $z_{{\it \Pi},{\rm max}}$ &  $S_{\it \Pi}$ & $W_{\it \Pi}$ & $\psi$ & shape \\
	  &[arcsec, kpc] &  [kpc]& [kpc] &  [kpc] & [kpc] & [kpc] &[$\degree$] & \\
	\hline \hline \\
	NGC~\, 128 ($a$)$^{\dagger}$ & -- & 0.042$\pm$0.003 & 0.72$\pm$0.02 & 0.70$\pm$0.02 & \: 1.68$\pm$0.59 & 0.53$\pm$0.06 & 43.49 & hump\\
	NGC~\, 128 ($b$)$^{\dagger}$ & -- & 0.091$\pm$0.004 & 3.44$\pm$0.02 & 2.58$\pm$0.02 & -- & -- & 36.29 & --\\
	NGC~\, 128 ($b$)$^{\ddagger}$& -- & 0.087$\pm$0.008 & 3.54$\pm$0.09 & 2.66$\pm$0.09 & 24.64$\pm$6.62 & 3.42$\pm$0.24 & 36.39 & hump\\
	NGC~\, 678 & 47.92$\pm$7.69, 9.25$\pm$1.48 & 0.054$\pm$0.006 & 2.67$\pm$0.11  & 2.30$\pm$0.11 & \: 5.51$\pm$2.78 & 1.27$\pm$0.31 & 40.72 & pyramid\\
	NGC~2549 ($a$)$^{\dagger}$ & 43.56$\pm$0.58, 4.14$\pm$0.06  & 0.037$\pm$0.002 & 0.53$\pm$0.01 & 0.50$\pm$0.01 & \: 0.96$\pm$0.06 & 0.33$\pm$0.02 & 42.35 & hump\\
	NGC~2549 ($b$)$^{\ddagger}$& 43.56$\pm$0.58, 4.14$\pm$0.06  & 0.030$\pm$0.002 & 1.57$\pm$0.03 & 1.12$\pm$0.03 & \: 1.80$\pm$0.60 & 0.76$\pm$0.08 & 34.53 & hump\\
	NGC~2654 & 31.52$\pm$0.36, 3.69$\pm$0.04 & 0.066$\pm$0.008 & 2.25$\pm$0.04 & 1.11$\pm$0.07 & \: 4.31$\pm$2.15 & 0.77$\pm$0.19 & 26.08 & HoP\\
	NGC~2683 & 87.18$\pm$4.5, \: 2.79$\pm$0.14  & 0.036$\pm$0.003 & 1.80$\pm$0.04 & 0.95$\pm$0.02 & \: 2.76$\pm$0.66 & 0.93$\pm$0.05 & 27.47 & saw-tooth\\ 
	NGC~3628 & 68.09$\pm$1.41, 2.93$\pm$0.06 & 0.055$\pm$0.007 & 3.08$\pm$0.03 & 1.13$\pm$0.03 & \: 6.74$\pm$1.21 & 1.79$\pm$0.07 & 20.05 & top-hat\\
	NGC~4111 & 30.51$\pm$0.34, 2.29$\pm$0.03 & 0.035$\pm$0.003 & 0.92$\pm$0.02 & 0.56$\pm$0.02 & \: 1.78$\pm$0.59 & 0.62$\pm$0.06 & 31.08 & hump\\
	NGC~4469 & 40.94$\pm$1.55, 0.78$\pm$0.03 & 0.067$\pm$0.007 & 0.69$\pm$0.01 & 0.38$\pm$0.01 & \: 1.36$\pm$0.34 & 0.26$\pm$0.02 & 27.78 & hump\\
	NGC~4710 & 31.05$\pm$0.25, 2.21$\pm$0.02 & 0.059$\pm$0.015 & 1.21$\pm$0.08 & 0.60$\pm$0.08 & \: 2.71$\pm$1.22 & 0.60$\pm$0.11 & 26.24 & HoP\\ 
	NGC~7332 & 20.30$\pm$0.15, 1.95$\pm$0.01 & 0.037$\pm$0.003 & 1.44$\pm$0.03 & 0.86$\pm$0.06 & \: 3.10$\pm$0.88 & 1.06$\pm$0.08 & 30.36 & hump\\
	ESO~443-042 & 36.13$\pm$1.12, 6.32$\pm$0.20 & 0.109$\pm$0.045 & 3.32$\pm$0.20 & 0.87$\pm$0.20 & \: 8.93$\pm$4.79 & 1.02$\pm$0.28 & 14.38 & HoP\\
	\hline
	$^{\dagger}$from $HST$ image,& $^{\ddagger}$from SDSS image
	\end{tabular}
}
\label{tab:sample}
\end{minipage}
\\
\begin{flushleft}
Note: Due to the non-exponential disc in NGC~128, we do not have an (exponential) scale length $h$ value.
\end{flushleft}
\end{table*}

\begin{table*}
\begin{minipage}{80mm}
\caption{The X/P Diagnostics, in units of $h$ and arcsec}
\centerline{
	\begin{tabular}{l c c c c}
	\hline 
	Galaxy  & $R_{{\it \Pi},{\rm max}}$ & $z_{{\it \Pi},{\rm max}}$ &  $S_{\it \Pi}$ & $W_{\it \Pi}$ \\
	  &  [units of $h$, arcsec] & [units of $h$, arcsec] &  [units of $h$, arcsec] & [units of $h$, arcsec] \\
	\hline \hline \\
	NGC~\, 128 ($a$)$^{\dagger}$    & 0.12$\pm$0.01, \: 2.50$\pm$0.15 & 0.12$\pm$0.01, \: 2.43$\pm$0.15 & \:\:\:\:\:\:\: -- \:\:\:\:\:\: , \:\:\: 5.8$\pm$2.1 \: & \:\:\:\:\:\:\: -- \:\:\:\:\:\: , \: 1.8$\pm$0.2\\
	NGC~\, 128 ($b$)$^{\dagger}$    & 0.59$\pm$0.01, 11.95$\pm$0.15 & 0.44$\pm$0.01, \: 8.97$\pm$0.15 & -- & --\\
	NGC~\, 128 ($b$)$^{\ddagger}$   & 0.61$\pm$0.03, 12.28$\pm$0.59 & 0.46$\pm$0.03, \: 9.25$\pm$0.59 & \:\:\:\:\:\:\: -- \:\:\:\:\:\: , \: 85.6$\pm$23.0 & \:\:\:\:\:\:\: -- \:\:\:\:\:\: , 11.9$\pm$0.9 \\
	NGC~\, 678   & 0.29$\pm$0.05, 13.83$\pm$1.13 & 0.25$\pm$0.09, 11.90$\pm$1.13 & 0.60$\pm$0.30, \: 28.6$\pm$14.4 & 0.14$\pm$0.03, \: 6.6$\pm$1.6\\ 
	NGC~2549 ($a$)$^{\dagger}$ & 0.13$\pm$0.04, \: 5.60$\pm$0.15 & 0.12$\pm$0.01, \: 5.29$\pm$0.15 & 0.23$\pm$0.06, \: 10.1$\pm$2.6 \: & 0.08$\pm$0.01, \: 3.5$\pm$0.2\\
	NGC~2549 ($b$)$^{\ddagger}$ & 0.38$\pm$0.02, 16.51$\pm$0.59 &0.27$\pm$0.02, 11.77$\pm$0.59 & 0.43$\pm$0.15, \: 18.9$\pm$6.3 \: & 0.19$\pm$0.02, \: 8.0$\pm$0.9\\
	NGC~2654 & 0.61$\pm$0.04, 19.19$\pm$1.13 & 0.30$\pm$0.04, \: 9.51$\pm$1.13	& 1.17$\pm$0.58, \: 36.8$\pm$18.4& 0.21$\pm$0.05, \: 6.6$\pm$1.6 \\
	NGC~2683  & 0.65$\pm$0.07, 56.32$\pm$1.13 & 0.34$\pm$0.07, 29.57$\pm$1.13 & 0.99$\pm$0.24, \: 86.1$\pm$20.7& 0.34$\pm$0.02, 29.2$\pm$1.6 \\ 
	NGC~3628   & 1.05$\pm$0.05, 71.69$\pm$1.13 & 0.38$\pm$0.05, 26.16$\pm$1.13 & 2.30$\pm$0.41, 156.7$\pm$28.1 & 0.61$\pm$0.02, 41.6$\pm$1.6 \\
	NGC~4111 & 0.40$\pm$0.02, 12.30$\pm$0.59 & 0.24$\pm$0.02, \: 7.41$\pm$0.59 & 0.78$\pm$0.26, \: 23.8$\pm$7.8 \: & 0.27$\pm$0.03, \: 8.3$\pm$0.9 \\
	NGC~4469   & 0.89$\pm$0.04, 36.32$\pm$0.59 & 0.48$\pm$0.07, 19.81$\pm$0.59 & 1.75$\pm$0.44, \: 71.6$\pm$18.1 & 0.33$\pm$0.02, 13.5$\pm$0.8 \\
	NGC~4710   & 0.55$\pm$0.03, 17.08$\pm$1.13 & 0.27$\pm$0.04, \: 8.42$\pm$1.13 & 1.23$\pm$0.55, \: 38.2$\pm$17.2 & 0.27$\pm$0.05, \: 8.5$\pm$1.6 \\
	NGC~7332  & 0.74$\pm$0.03,  15.00$\pm$0.59 & 0.44$\pm$0.03, \: 8.92$\pm$0.59 & 1.59$\pm$0.45, \: 32.2$\pm$9.2 \: & 0.55$\pm$0.04, 11.1$\pm$0.8 \\
	ESO~443-042  & 0.55$\pm$0.04,  19.96$\pm$1.13 & 0.14$\pm$0.04, \: 5.11$\pm$1.13 & 1.41$\pm$0.76, \: 51.1$\pm$27.4 & 0.16$\pm$0.04, \: 5.9$\pm$1.6\\
	\hline 
	$^{\dagger}$from $HST$ image,& $^{\ddagger}$from SDSS image
	\end{tabular}
}
\label{tab:sample_arcsec-h}
\end{minipage}
\end{table*}

\begin{enumerate}

\item NGC~2549 -- This is the second galaxy where we detect both an inner and an outer X/P structure. As for NGC~128, we denote the inner peanut with $(a)$ and the outer with $(b)$. Also similarly to NGC~128, we measured the diagnostics for $(a)$ from an $HST$ image, this time taken with the Wide Field (WF) + Planetary Camera (PC), WFPC 2. We refer the reader to Fig. 8 in C15 to see the image, reconstruction and residual, and their Fig. 10 to see the $B_6$ profile. However, because the peak in the outer peanut coincided spatially with the gap between the WF and PC chips, we again opted to model peanut $(b)$ using the SDSS $r$--band image. We show the latter in Figure \ref{fig:sample-1} with an increased number of contours overplotted, to aid the eye in making out the inner peanut. Although we also see the signature of peanut $(a)$ in the SDSS $B_6$ profile, we prefer the higher spatial resolution (and lower seeing) of the $HST$ image for measuring it, since poorer seeing smears out and systematically underestimates the X/P peak (we see this in Table \ref{tab:sample}, where the $B_6$ profile of NGC~128 has lower amplitude for both peanuts in the SDSS data compared to the $HST$ data). Again, we indicate the major-axis radii of peanuts ($a$) and ($b$) on the surface brightness profile decomposition (Figure \ref{fig:n2549_decomp}), and again we see that both radii match approximately the ends of the two bars: $\sim 10\arcsec$ for bar ($a$) (in good agreement with the range $8-10\arcsec$ reported in \citealt{LaurikainenEA2011}) and $\sim 30\arcsec$ for bar ($b$).

\item NGC~2654 and NGC~2683 -- Both of these galaxies display regions of comparatively low brightness, bordered by `spurs', along the major axis on either side of the `bulge', suggesting that they are boxy/peanut (\citealt{ErwinDebattista2013}) or barlens galaxies (\citealt{LaurikainenEA2014}, \citealt{AthanassoulaEA2015}) viewed close to, but not exactly, edge-on, with a thin bar aligned such that its viewing angle $\alpha < 90\degree$ (i.e., not exactly side-on).

\item NGC~4111 -- This galaxy $appears$ to host an inner peanut as well, at $\sim 2\arcsec$, in the SDSS $i$--band image. However, a closer inspection reveals an edge-on dust torus perpendicular to the disc plane at precisely this spatial scale. The obscuration which this induces causes the isophotes to appear `pinched', just as they would if an inner peanut were there. Therefore, we chose to not classify this as an inner peanut, at least not until higher-resolution NIR imaging is available (NB: The `inner peanut' does not show at all in the $B_6$ profile extracted from the $Spitzer$ imaging, both 3.6$\mu$ and 4.5$\mu$).

\item NGC~4469 -- The major axis of this galaxy is heavily obscured by dust. As there is no NIR image available, we used the reddest SDSS passband ($z$) image, which has relatively poor S/N, and still shows the dust slightly, in the inner regions (see residual map). Nevertheless, the $B_6$ profile, though noisy, is robust and very nicely shows the peanut signature.

\item NGC~4710 -- The image of this galaxy (and, much more obviously, the residual image) shows two pronounced stripes, which were attributed to dark stripes on the CCDs caused by a bright star (there are two because the image is a mosaic of different individual exposures).

\end{enumerate}

For each galaxy in the sample, we extracted the major-axis surface brightness profile, $\mu(R_{\rm maj})$, and decomposed it with {\sc profiler} to obtain the disc scale length. As we have discussed above, X/P structures are closely associated with the galactic discs in which they are embedded. If a disc buckles and gives rise to a bar, which also buckles and gives rise to a peanut, the peanut may reflect the disc's fundamental properties. A pertinent question, therefore, is: are the peanut's metrics (size, shape) random, or do they depend on their host disc's properties? We address this question by measuring the disc's scale length $h$, and expressing the peanut height, projected length, peak width and integrated strength in units of $h$, as well in kiloparsecs. Because our test-case galaxy, NGC~128, displays very unusual (warped) morphology, and the surface brightness profile of its disc is not exponential, but S\'ersic ($n \sim 0.4$; see Figure \ref{fig:n128_decomp}), it is not meaningful to assign it a scale length. As such, we only report values of $h$ for the rest of the galaxies in our sample.

The PSF profiles used to convolve the model $\mu(R_{\rm maj})$ were always characterised from bright stars in the image of each galaxy. A second example of a {\sc profiler} decomposition is shown for NGC~2549 in the top panel of Figure \ref{fig:n2549_decomp}, where the model consisted of a bulge (S\'ersic function, red curve), a disc (exponential, blue curve) and a bar associated with each of the two peanuts (Ferrers functions -- \citealt{Ferrers1877}, cyan curves), once again nicely confirming the bar-peanut connection in this second galaxy with a nested peanut structure. 

\subsection{Scaling Relations}\label{sec:analysis-relations}

The data shows a remarkably tight correlation between $R_{{\it \Pi},{\rm max}}$ and the integrated peanut strength $S_{\it \Pi}$ (i.e., the area under the $B_6$ curve). We display this relationship in the upper panels of Figure \ref{fig:RZ-S}. As $B_6$ is dimensionless, with any radial dependence normalised out through Equation \ref{equ:normalisation}, it is independent of $R_{{\it \Pi},{\rm max}}$, and therefore so is $S_{\it \Pi}$, which is derived from the profile. This relation implies that stronger peanuts are located further out in a galaxy, and it is further surprising that, while reasonably tight in units of kpc, it also holds when expressing both quantities in units of the disc scale length $h$ (Figure \ref{fig:RZ-S}, top-left panel). This suggests that when a disc defined by a scale length $h$ buckles and forms a bar, which in turn gives rise to a peanut, the resulting peanut still retains the information about its parent disc's scale length. This is consistent with the scenario proposed by \cite{BureauEA2006}, who suggest that the different components of galaxies with X/P features (including the peanut component itself) are dynamically coupled, and most likely all originated from the disc material.

\begin{figure*}
\centering
\begin{center}$
\begin{array}{cc}
\includegraphics[width=.475\textwidth]{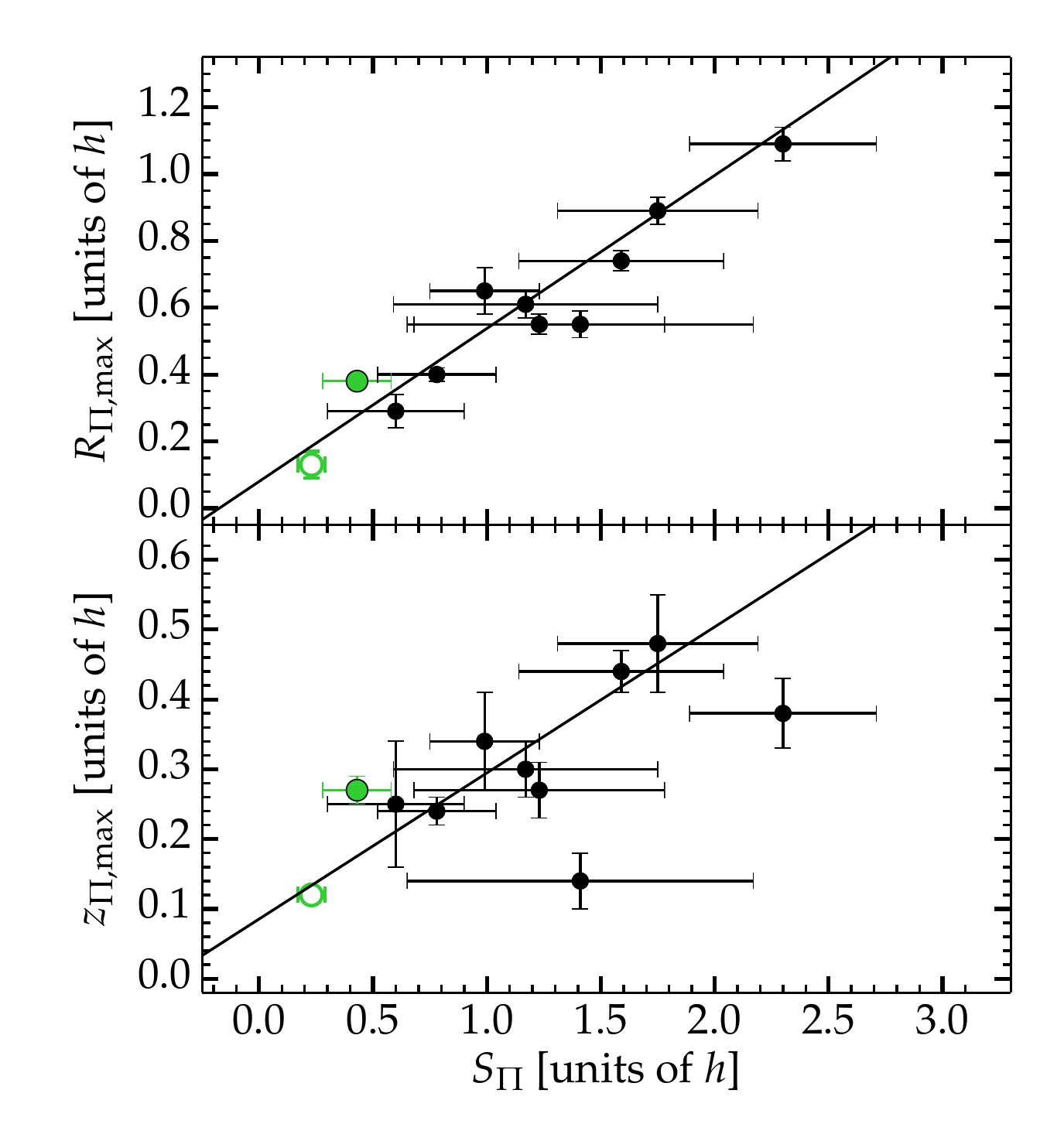} & \includegraphics[width=.475\textwidth]{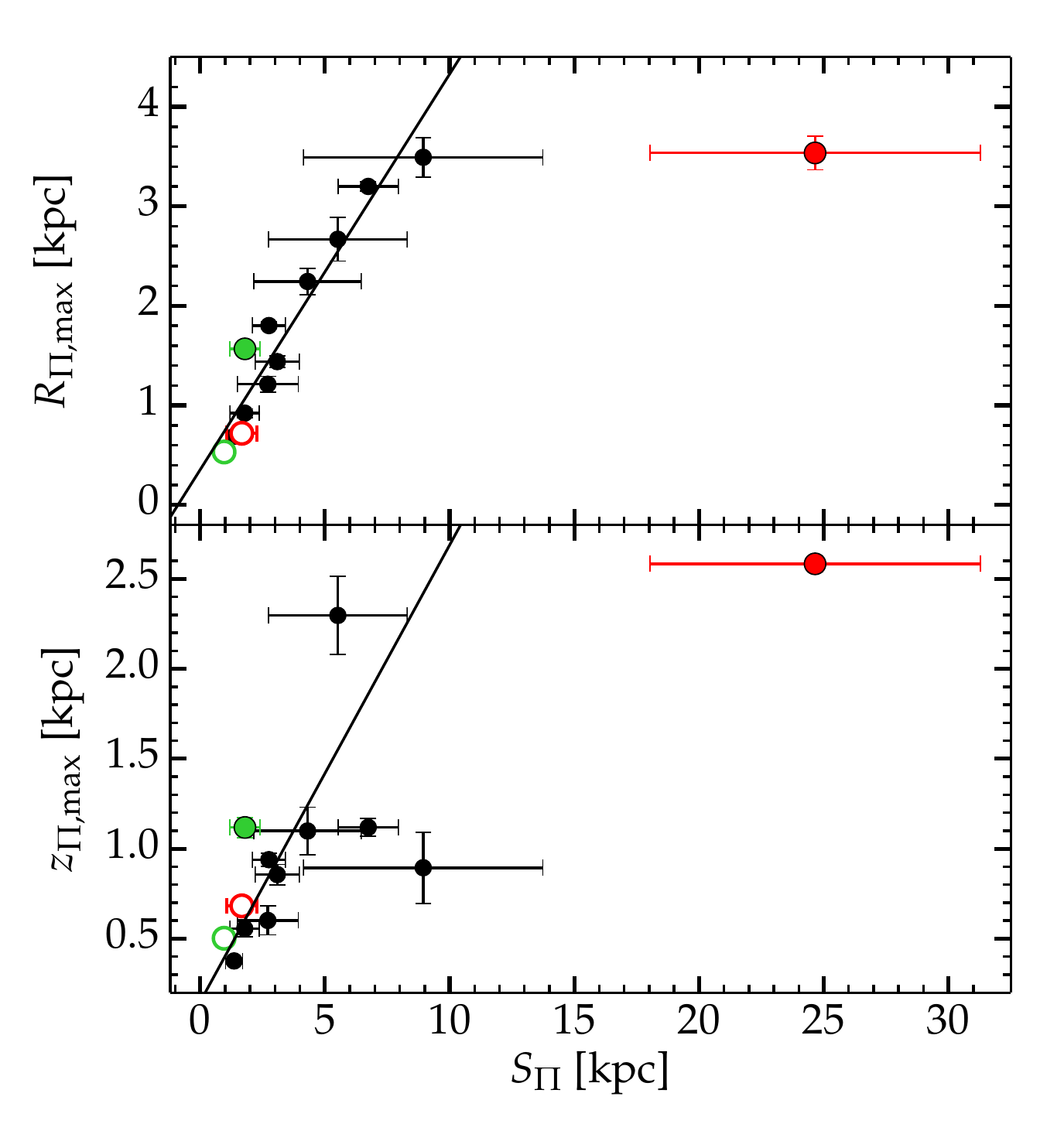}
\end{array}$
\end{center}
\caption{{\it Left}: Projected radial length (top), and vertical height above the disc plane (bottom) of the X/P structures, as a function of integrated strength $S_{\Pi}$, all in units of the disc scale length $h$. {\it Right}: The same quantities as in the left-hand panels, but plotted in kpc. Red symbols correspond to NGC~128 while green symbols correspond to NGC~2549; filled circles correspond to outer peanuts whereas open circles to inner peanuts. The linear relations shown are given by Equations \ref{equ:relRkpc} -- \ref{equ:relzh}, and exclude the outer, outlying peanut of NGC~128.}
\label{fig:RZ-S}
\end{figure*}

We performed basic linear fits to all the trends plotted in Figure \ref{fig:RZ-S}, which we show as Equations \ref{equ:relRkpc}, \ref{equ:relRh}, \ref{equ:relzkpc} and \ref{equ:relzh}. As NGC~128 ($b$) appears to be an outlier from all the trends, we excluded it from the fits. The fact that this galaxy is an outlier is not entirely surprising considering its peculiarity. More specifically, in addition to its visibly distorted morphology, there is also evidence that it hosts a counter-rotating gas disc, tilted at $\sim 26\degree$ from its major axis (\citealt{EmsellemArsenault1997}). Perhaps NGC~128 is an example of the scenario proposed by \cite{BinneyPetrou1985}, who argue that galaxy interactions (slow accretion: galactic cannibalism) may if not generate, then at least enhance the peanut shape. NGC~128 is just such a galaxy, and we deliberately show its companion NGC~127 in Figure \ref{fig:n128_SDSS}, which visibly exchanges material with it. In fact, several of our galaxies display somewhat distorted ($S$-shaped) disc planes, possibly indicating instability due to tidal interaction (e.g., NGC~7332). Note that we expect non-symmetric harmonics (both $A_n$ and $B_n$ coefficients) in galaxies observed in the process of bar-buckling. For such cases, the dominant shape of the instability would be a strong banana shape, which would likely be described by, e.g., $n = 3$ or, if the disc is comparatively bright at those radial scales, by $n=5$ (in edge-on disc projection, side-on bar projection).

All the linear regressions were performed with the bisector method. The fit uncertainties were computed through bootstrap re-sampling (10\,000 samples), which is best suited for such sparse data.

\begin{equation}
\label{equ:relRkpc}
\frac{R_{{\it \Pi},{\rm max}}}{\rm kpc} = (0.03\pm0.08) + (0.53\pm0.07) \frac{S_{\it \Pi}}{\rm kpc}
\end{equation}

\begin{equation}
\label{equ:relRh}
\frac{R_{{\it \Pi},{\rm max}}}{h} = (0.08\pm0.07) + (0.46\pm0.05) \frac{S_{\it \Pi}}{h}
\end{equation}

\begin{equation}
\label{equ:relzkpc}
\frac{z_{{\it \Pi},{\rm max}}}{\rm kpc} =  (0.15\pm0.17) + (0.25\pm0.09) \frac{S_{\it \Pi}}{\rm kpc}
\end{equation}

\begin{equation}
\label{equ:relzh}
\frac{z_{{\it \Pi},{\rm max}}}{h} = (0.09\pm0.06) + (0.21\pm0.05) \frac{S_{\it \Pi}}{h}
\end{equation}\\

That the peanut strength correlates with $both$ its height and its length is not surprising given that the latter two appear to be correlated to each other.  As shown in Figure \ref{fig:R-Z}, the shorter ($R_{{\it \Pi},{\rm max}} \lesssim  2.5$ kpc) X/P structures follow a characteristic height-to-length ratio of $\sim 0.5$ - 0.6, which breaks down for the longer X/P structures. While the bar orientation (the sin($\alpha$) term mentioned prior to Equation \ref{equ:r_pi}) may impact this trend, driving down the observed projected length if not viewed perfectly side-on, the discrepancies can be perhaps also understood from the point of view of the peanut's age or additional, non-secular processes, as we will speculate in Section \ref{sec:discussion}.

\begin{figure}
\centering
\begin{center}$
\begin{array}{cc}
\includegraphics[width=.99\columnwidth]{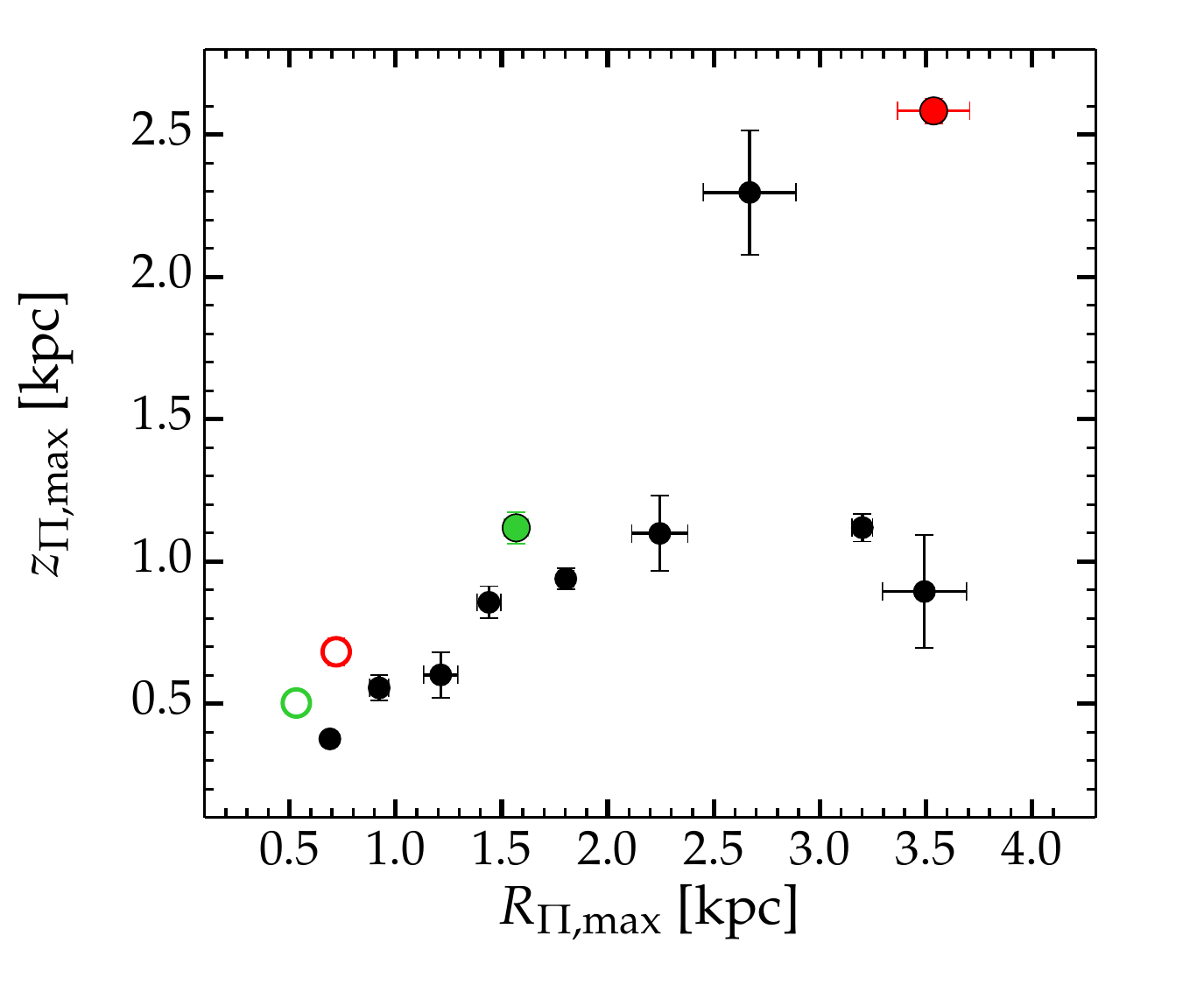}
\end{array}$
\end{center}
\caption{The smaller peanuts appear to follow a characteristic ratio of $z_{{\it \Pi}{\rm , max}}$/$R_{{\it \Pi},{\rm max}}$ of $ \sim 0.5 - 0.6$, while the larger peanuts appear to deviate from this. As in Figure \ref{fig:RZ-S}, the green symbols correspond to NGC~2549, while the red symbols correspond to NGC~128, and filled circles correspond to outer peanus whereas open circles to inner peanuts.}
\label{fig:R-Z}
\end{figure}

In terms of the peanut height above the disc ($z_{{\it \Pi},{\rm max}}$), we found that for all galaxies in our sample this lies in the range $0.1h \leq z_{\it \Pi} \leq 0.5h$, or equivalently, within $\sim 2.6$ kpc. The stronger peanuts also seem to reach greater heights above the disc plane, though this trend is not as tight as the location--strength relation. 

\subsubsection{Peanut--Disc Scaling Relations}

Just as `bar strength' has been tested for correlations with properties of the host disc, such as star formation (e.g., \citealt{MartinetFriedli1997}; \citealt{Aguerri1999}), nuclear activity (e.g., \citealt{LaurikainenSaloRautiainen2002}; \citealt{LaurikainenSaloButa2004}; \citealt{CisternasEA2013}), central velocity dispersion (\citealt{DasEA2008}), various gaseous features (e.g., \citealt{PeeplesMartini2006}; \citealt{KimSeoKim2012}; see also \citealt{AthanassoulaEA2013}), etc., we can now test for correlations between peanut strength and the physical characteristics of the host disc.  It is hoped that this will provide a further quantitative setting for testing the different mechanisms proposed for peanut formation, and thus a deeper understanding of their evolutionary path.

We observed a positive, though weak, correlation of the galaxy's $v_{\rm rot}/\sigma_{\star}$ (Figure \ref{fig:V/Sig_S-R}). This trend shows, in essence, that peanuts are more pronounced in faster rotating galaxies, once more pointing towards their link with the host disc. As before, we performed linear (bisector method) fits to the relations in Figure \ref{fig:V/Sig_S-R}, and estimated the uncertainties via bootstrap re-sampling. We show the relations as Equations \ref{equ:relvsRkpc} and \ref{equ:relvsSh}. We note, however, the exclusion of 5 data points from these fits, namely NCG~2654 and ESO~443-042, as we lacked values for their velocity dispersion, $\sigma_{\star}$; NGC~128($a$) and ($b$), because of ambiguities arising from this galaxy's complicated morphology (see Section \ref{sec:discussion}); and NGC~4469, as we find that its reported value of $v_{\rm rot} = 18\pm9$ km\,s$^{-1}$ is implausibly low for a rotation-supported, edge-on disc galaxy. 
  
\begin{figure*}
\centering
\begin{center}$
\begin{array}{cc}
\includegraphics[width=.9\textwidth]{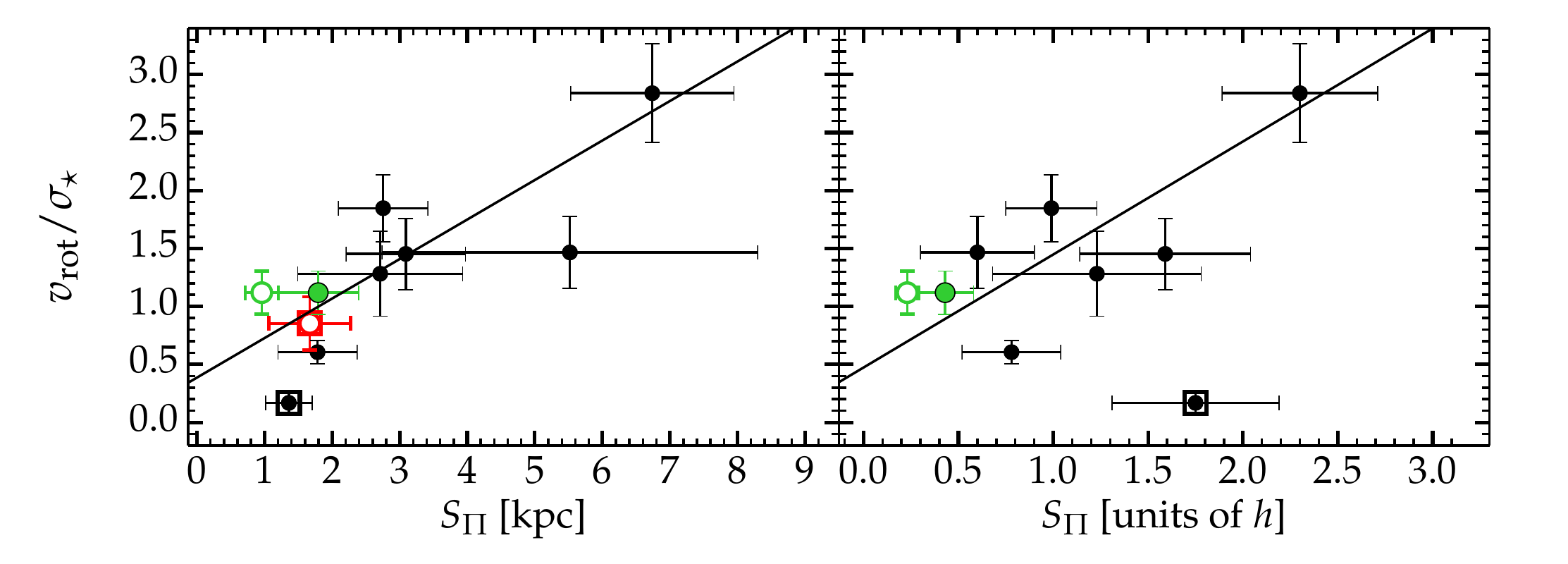} & \\ \includegraphics[width=.9\textwidth]{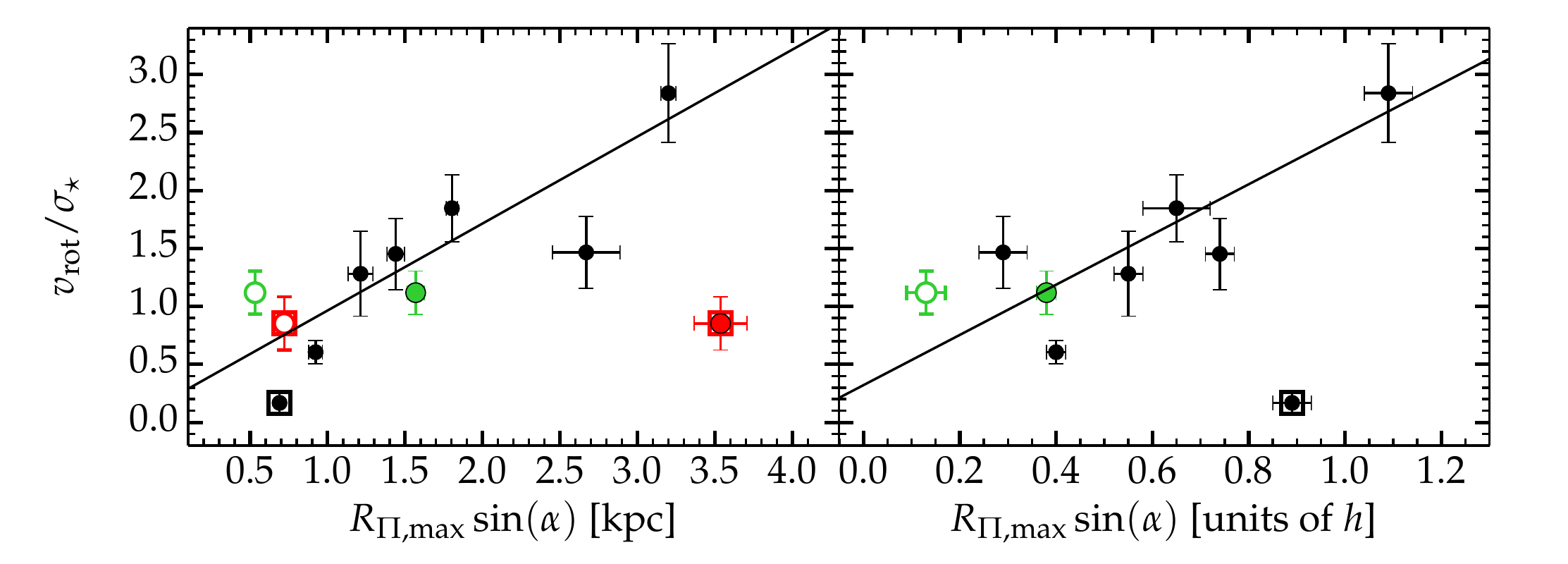}
\end{array}$
\end{center}
\caption{{\it Top}: The $v/\sigma$ ratio as a function of peanut strength. The colour scheme is analogous to Figures \ref{fig:RZ-S} and \ref{fig:R-Z}. Data points enclosed in squares were excluded from the linear fits (see Section \ref{sec:analysis-relations}), which are given by Equations \ref{equ:relvsRkpc} and \ref{equ:relvsSh}).The data point corresponding to the outer peanut of NGC~128 is a significant outlier of this trend, and is outside (to the right of) the plot area (it was excluded from the fits, where applicable). {\it Bottom}: The $v/\sigma$ ratio as a function of peanut length.}
\label{fig:V/Sig_S-R}
\end{figure*}

\begin{equation}
\begin{split}
\label{equ:relvsRkpc}
\frac{v_{\rm rot,\,gas}}{\sigma_{\star}} & = (0.21\pm0.30) +   (0.75\pm0.16) \frac{R_{{\it \Pi},{\rm max}}}{\rm kpc}\\
& = (0.32\pm0.48) +   (2.17\pm0.82) \frac{R_{{\it \Pi},{\rm max}}}{h}
\end{split}
\end{equation}

\begin{equation}
\begin{split}
\label{equ:relvsSh}
\frac{v_{\rm rot,\,gas}}{\sigma_{\star}} & = (0.39\pm0.25) +   (0.34\pm0.06) \frac{S_{\it \Pi}}{\rm kpc} \\
& = (0.47\pm0.28) +   (0.98\pm0.18) \frac{S_{\it \Pi}}{h}
\end{split}
\end{equation}

We did not find any correlation between any of our X/P diagnostics and the galaxy global, $K_s$--band magnitude.

\section{Discussion}\label{sec:discussion}

Having diagnosed and quantified the X/P features in our galaxy sample, we now shift the focus of the paper to discussing the information encoded in our peanut diagnostics. 

Most studies in the literature invoke either the buckling of bars (\citealt{RahaEA1991}) or vertical Lindblad Resonances (\citealt {CombesEA1990}) to describe essentially the same class of objects. While numerical simulations have been resoundingly successful at reproducing, visually, X/P-like structures, direct comparisons between observations and simulations have seldom been performed. As such, it remains an open question which physical mechanism is responsible for which type of X/P structure. Our study aims, among other things, to provide a set of useful measurements from real X/P galaxies which can act as constraints for $N$-body simulations of galactic dynamics (e.g., \citealt{SahaGerhard2013}).

The first point to note is that our sample galaxies are remarkably heterogeneous in terms of their radial $B_6$ profile shape. While all show an unambiguous peak in $B_6$, which is the mark of the peanut, the peak can be shaped like a $hump$ (e.g., NGC~128$a$, NGC~2549$a$ and $b$, NGC~4111), {\it hump-on-plateau} (NGC~128$b$, NGC~2654, NGC~4710, ESO~443-042), {\it top-hat} (NGC~3628), {\it saw-tooth} (steady rise followed by sharp decline; NGC~2683, NGC~4469) or even $pyramid$ (NGC~678). This is not very surprising since bars (and triaxial ellipsoids in general) can host a large variety of orbit families (\citealt{PatsisSkokosAthanassoula2002}, \citealt{PatsisKatsanikas2014}, \citealt{PatsisKatsanikas2014b}, \citealt{ValluriEA2015}), each of which potentially leaving its characteristic imprint on the photometric morphology of the host galaxy.

If a bar buckles, it forms a peanut in the inner regions (\citealt{RahaEA1991}, \citealt{AthanassoulaMV2009}), and one would expect a relatively flat $B_6$ profile (e.g., top-hat) along the entire peanut length. On the other hand, a resonance mechanism usually occurs in a narrow radial range, thus corresponding to a comparatively sharper peanut peak (e.g., hump or pyramid). It would be interesting to see from simulations $i)$ whether there is a characteristic $B_6$ shape for each of the two phenomenologies and $ii)$ how the shape evolves with time. 

We might be observing peanuts at different stages of their lifetime, i.e., newly-formed or old. As these features have been shown to drift outwards with time (\citealt{QuillenEA2014}), the radial length $R_{{\it \Pi},{\rm max}}$, coupled with the peak width, $W_{\it \Pi}$, of the $B_6$ profile, might be an indicator of their age. The latter point is supported by noticing that for the two galaxies which host nested peanuts (NGC~128 and NGC~2549), the inner peanut (which is presumably the younger) has a narrower span than the outer, presumably older X/P structure. For this work, however, we are limited by projection effects and can only measure the projected peanut length,  $R_{{\it \Pi},{\rm max}} = l_{{\it \Pi},{\rm max}}\,{\rm sin}(\alpha)$. This aspect is a strong restriction to any conclusions we may draw from Figures \ref{fig:RZ-S} to \ref{fig:V/Sig_S-R}. For now we speculate that we may see tentative evidence of radial drift (starting from the `characteristic ratio' and moving out in radius while keeping the same height) or length/height enhancements due to external, non-secular processes, such as tidal interactions), though we require additional information on the bar's orientation in the disc plane to draw any conclusions. The key to constraining $\alpha$ may lie in applying our method to edge-on (disc) projections of simulated galaxies viewed at different bar angles $\alpha$, ranging from side-on to end-on. The $B_6$ profiles, which may contain information about $\alpha$, could easily be recovered from isodensity contours of the simulation projections (rather than isophotes). Such a study excedes the scope of this paper, however, and we defer it for future works. 

\section{Conclusions}\label{sec:conclusions}

In this work we define five quantitative diagnostics of X/P structures in edge-on galaxies, based of the sixth Fourier mode ($B_6$) of their isophotes: $i)$ the peak amplitude of the $B_6$ radial profile, ${\it \Pi}_{\rm max}$, $ii)$ its projected length along the major-axis, $R_{{\it \Pi},{\rm max}}$, $iii)$ its height above the disc plane, $z_{{\it \Pi},{\rm max}}$,  $iv)$ the integrated strength, $S_{\it \Pi}$ (Equation \ref{equ:sharpness}) and $v)$ the width of the $B_6$ peak, $W_{\it \Pi}$. Additionally, we introduce a qualitative classification of X/P galaxies, based on the shape of the $B_6$ profile. 

We demonstrate our methodology on NGC~128, a galaxy with a very strong peanut, and extend our analysis to a sample of eleven other galaxies known to host such structures. This technique is accurate, easy to implemented and automate, and it performs best when using imaging with low dust obscuration (dust-free galaxies or NIR wavelegths). 

Our main findings can be summarised as follows:

\begin{itemize}

\item The $n=4$ Fourier harmonic of isophotes ($B_4$) does not describe the X/peanut--shaped structure. Out of all the Fourier harmonics tested ($0 \leq n \leq 10$), it is the $n=6$ order ($B_6$ term) which captures the peanut.

\item We detect, for the first time, nested peanuts (one inner and one outer) in two of the eleven galaxies of our sample, namely NGC~128 and NGC~2549.

\item The galaxies in our sample are quite heterogeneous in terms of their $B_6$ profile shapes (which range from hump-shaped, hump-on-plateau, top-hat, saw-tooth and pyramid, in our classification scheme). We speculate that these may provide insight into disentangling between the various peanut formation scenarios in the literature.

\item We identified trends between peanut projected length and strength, and between peanut height and strength. The stronger peanuts are located at larger radii and reach greater heights above the disc plane. These trends hold when expressed in units of kpc and  disc scale length, indicating that peanuts `know' about the disc in which they live. Together with an apparently characteristic height-to-length ratio for small peanuts, this constitutes valuable constraints for simulations.

\item We additionally identified a positive, though weak, correlation between peanut parameters (length and strength) and the galaxy's $v_{\rm rot}/\sigma_{\star}$, such that faster-rotating galaxies tend to host larger and more pronounced X/P structures. This provides yet more support for the peanut--host disc link, but would benefit from more data.

\end{itemize}

There are many catalogues of edge-on disc galaxies (e.g., \citealt{DalcantonBernstein2002}; \citealt{KregelEA2005}; \citealt{KautschEA2006}; \citealt{YoachimDalcanton2006}; \citealt{ComeronEA2011}; \citealt{BizyaevEA2014}) including even the late-type ultra-flat galaxies (\citealt{KarachentsevaEA2016}) that can now be quantitatively analysed for the presence of X-shaped features. {\sc Isofit} is also well placed to both search for and quantify, in addition to cataloguing, banana-shaped (in projection) bars in the act of buckling in real galaxies, through non-symmetric harmonic terms. 

Additionally, kinematic follow-up of nested peanuts may be insightful, and several integral field spectrographs such as the {\it Calar Alto Legacy Integral Field Area} (CALIFA; \citealt{SanchezEA2012}); the {\it Sydney-AAO Multi-object Integral field spectrograph} (SAMI; \citealt{CroomEA2012}),  the {\it Mapping Nearby Galaxies} survey (MaNGA; \citealt{BundyEA2015}), or the {\it Multi-Unit Spectroscopic Explorer} (MUSE; \citealt{BaconEA2010}) are ideal for this task (see for example \citealt{GonzalezEA2016}, who study the X/P galaxy NGC~4710 with the MUSE instrument). 

Furthermore, this method allows for direct comparisons between real, observed galaxies and simulations. Used in conjunction, an observational approach, coupled with $N$-body simulations, have the potential to disentangle the various X/P formation mechanisms proposed in the literature. 

Such studies, however, are beyond the scope of this paper and, along with a study of the Milky Way's own peanut--shaped bulge, we defer them for future works.

\section{Acknowledgements}

BCC expresses warm thanks to Fran\c{c}oise Combes for a very stimulating discussion on the topic of X/P galaxies. This research has made use of the NASA/IPAC {\it Extragalactic Database} (NED), and the NASA/ IPAC {\it Infrared Science Archive} (IRSA),
which are operated by the Jet Propulsion Laboratory, California Institute of Technology,
under contract with the National Aeronautics and Space Administration. Funding for SDSS-III has been provided by the Alfred P. Sloan Foundation, the Participating Institutions, the National Science Foundation, and the U.S. Department of Energy Office of Science. Part of this work is based on observations made with the NASA/ESA {\it Hubble Space Telescope}, and obtained from the {\it Hubble Legacy Archive}, which is a collaboration between the {\it Space Telescope Science Institute} (STScI/NASA), the {\it Space Telescope European Coordinating Facility} (ST-ECF/ESA) and the {\it Canadian Astronomy Data Centre} (CADC/NRC/CSA). We acknowledge the usage of the HyperLeda database.

\bibliography{references}
\bibliographystyle{natbib}

\appendix

\setcounter{figure}{0}
\renewcommand{\thefigure}{A\arabic{figure}}

\section{Modelling the Galaxies}

In this section we show the results of our \ifit/\cmo\ analysis for each of the galaxies in our sample. Specifically, in each panel of Figure \ref{fig:sample-1}, we show the galaxy image and its orientation, the image reconstruction made with \cmo, the residual image obtained by subtracting the reconstruction from the original image and, at the bottom, the $B_6$ profile as a function of peanut radius (see Figure \ref{fig:peanut_radius}). The image and its reconstruction both have their respective isophote contours overlayed (at identical levels), and the thicker contour corresponds to the maximum $B_6$ amplitude (${\it \Pi_{\rm max}}$).

\begin{figure*}
\centering
\caption{Each quadrant, from top to bottom panel: image, model, residual and $B_6$ profile. Thick contours corespond to ${\it \Pi_{\rm max}}$ (the isophote with maximal $B_6$ amplitude).}
\begin{center}$
\begin{array}{cc}
\includegraphics[width=.455\textwidth]{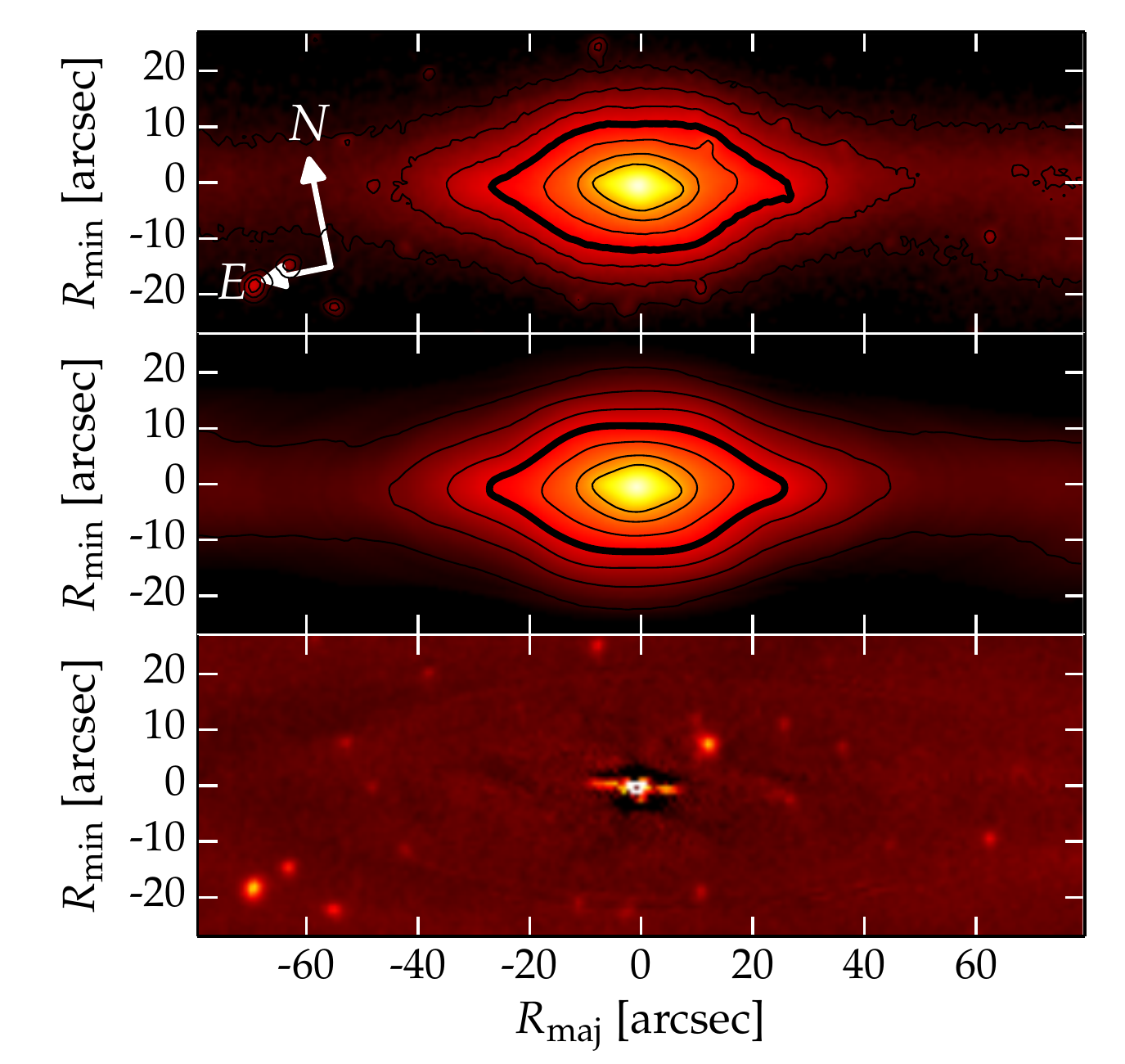} \vline & \includegraphics[width=.455\textwidth]{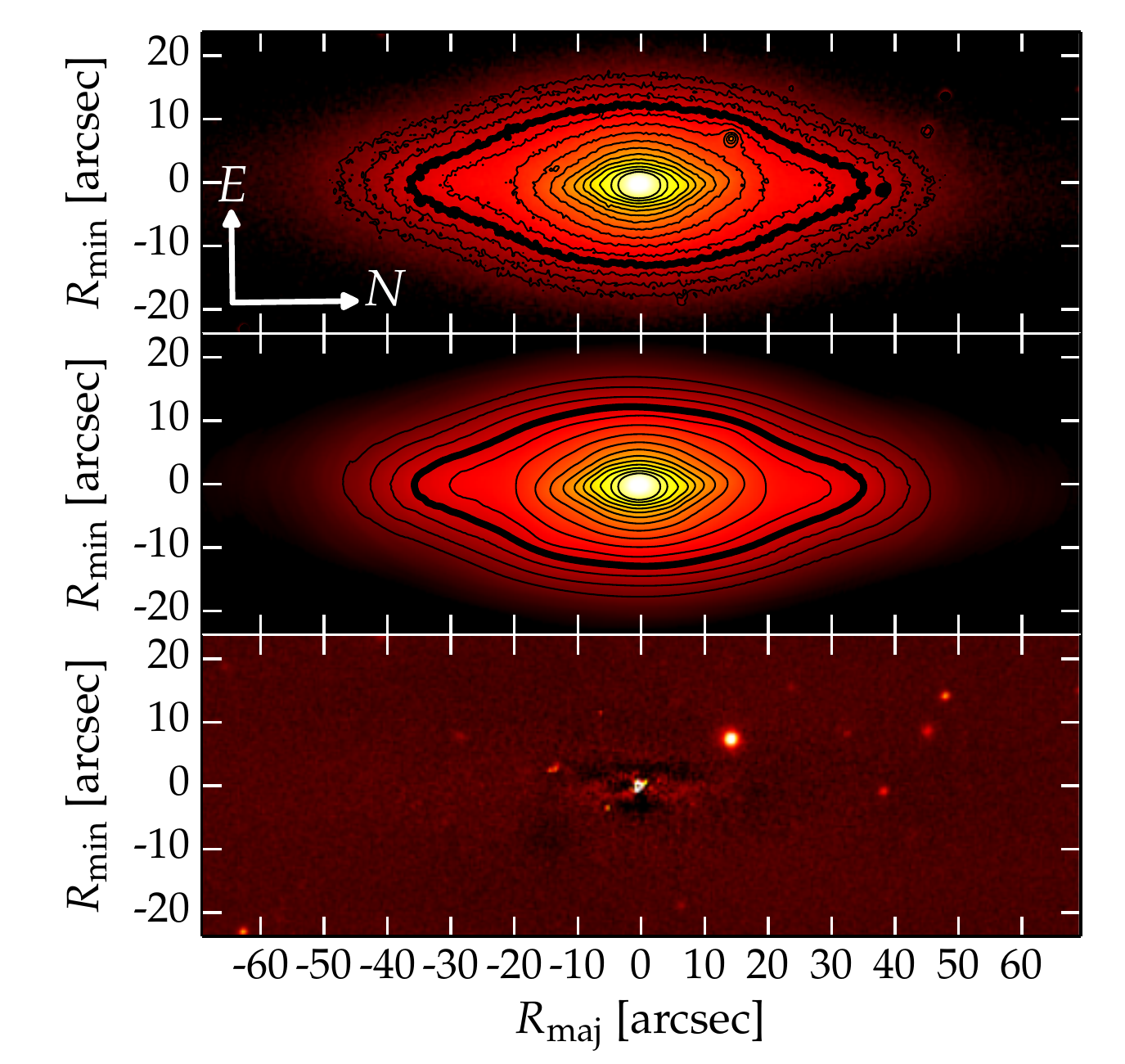} \\
\includegraphics[width=.455\textwidth]{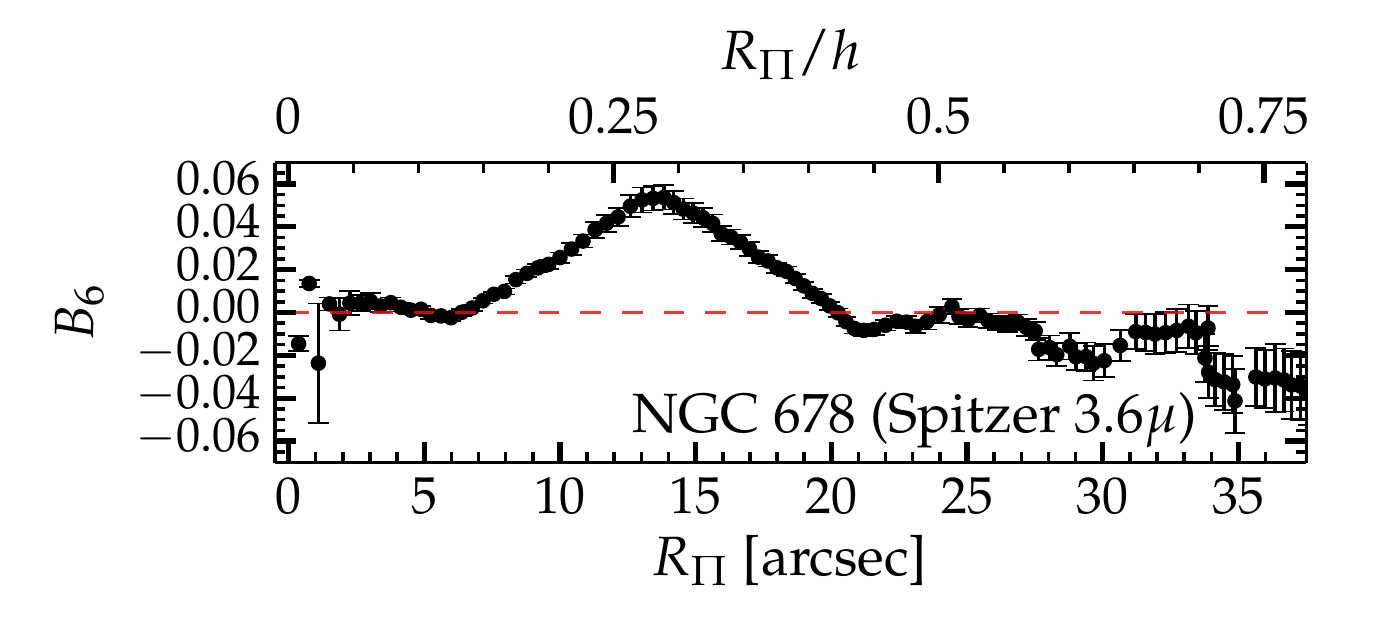} \vline & \includegraphics[width=.455\textwidth]{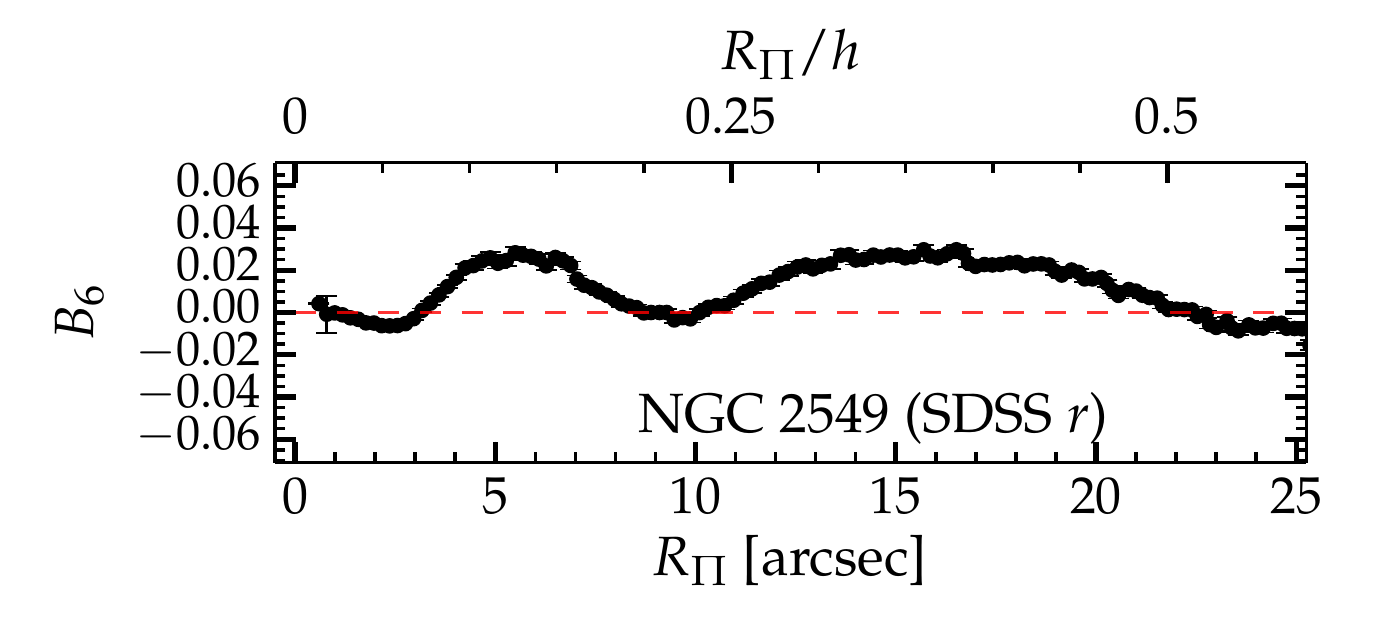}\\
\hline
\includegraphics[width=.455\textwidth]{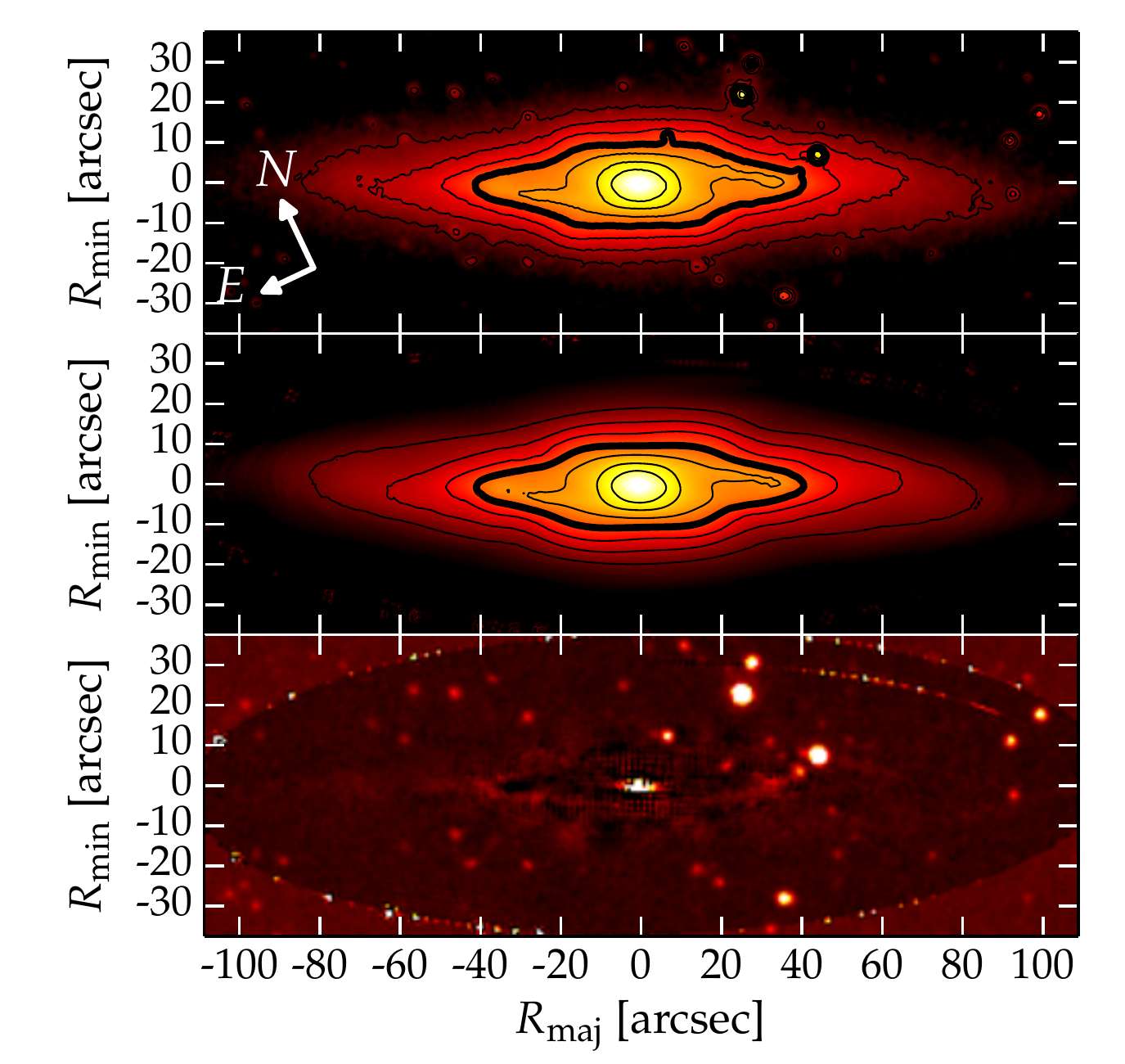} \vline & \includegraphics[width=.455\textwidth]{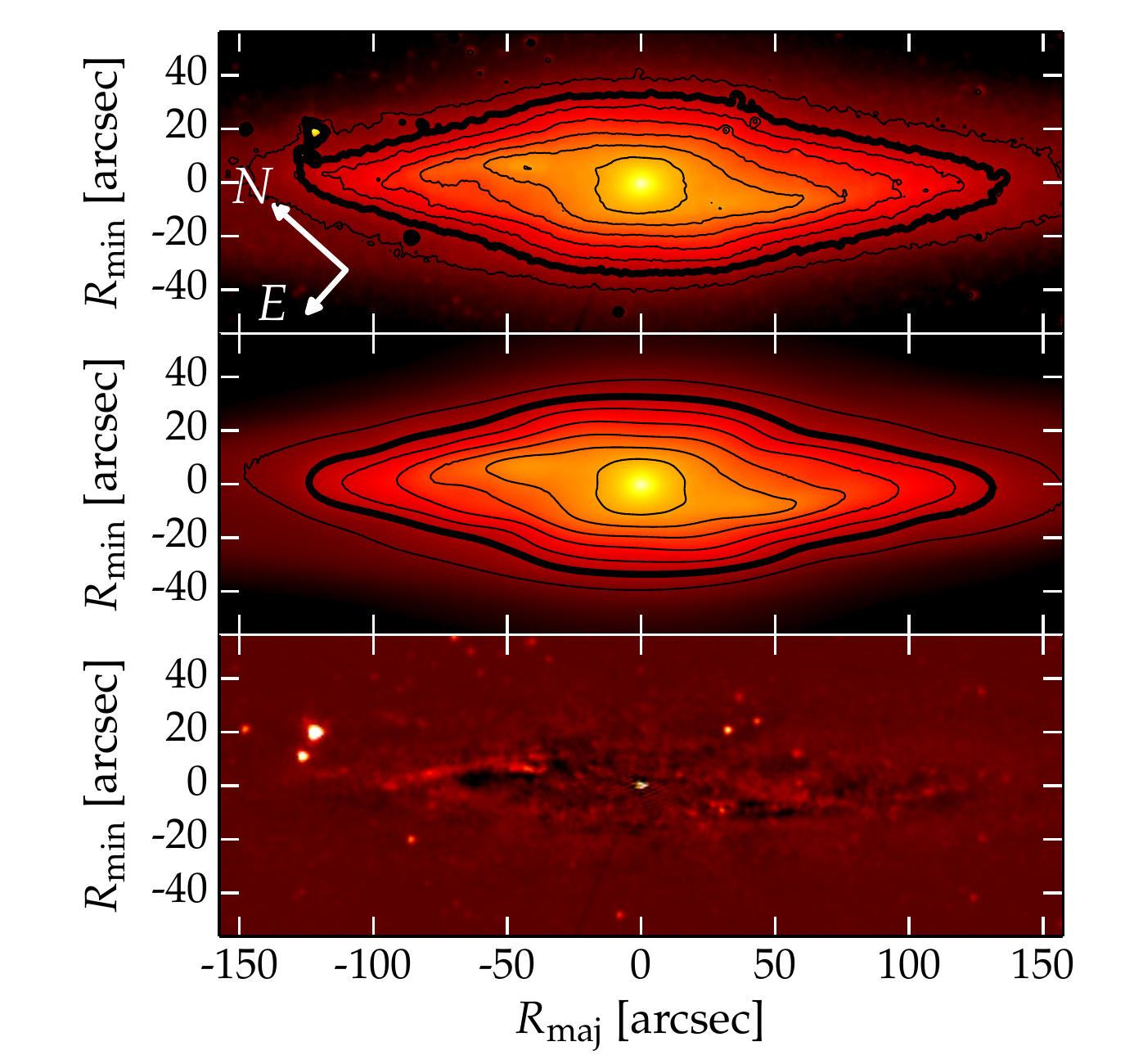} \\
\includegraphics[width=.455\textwidth]{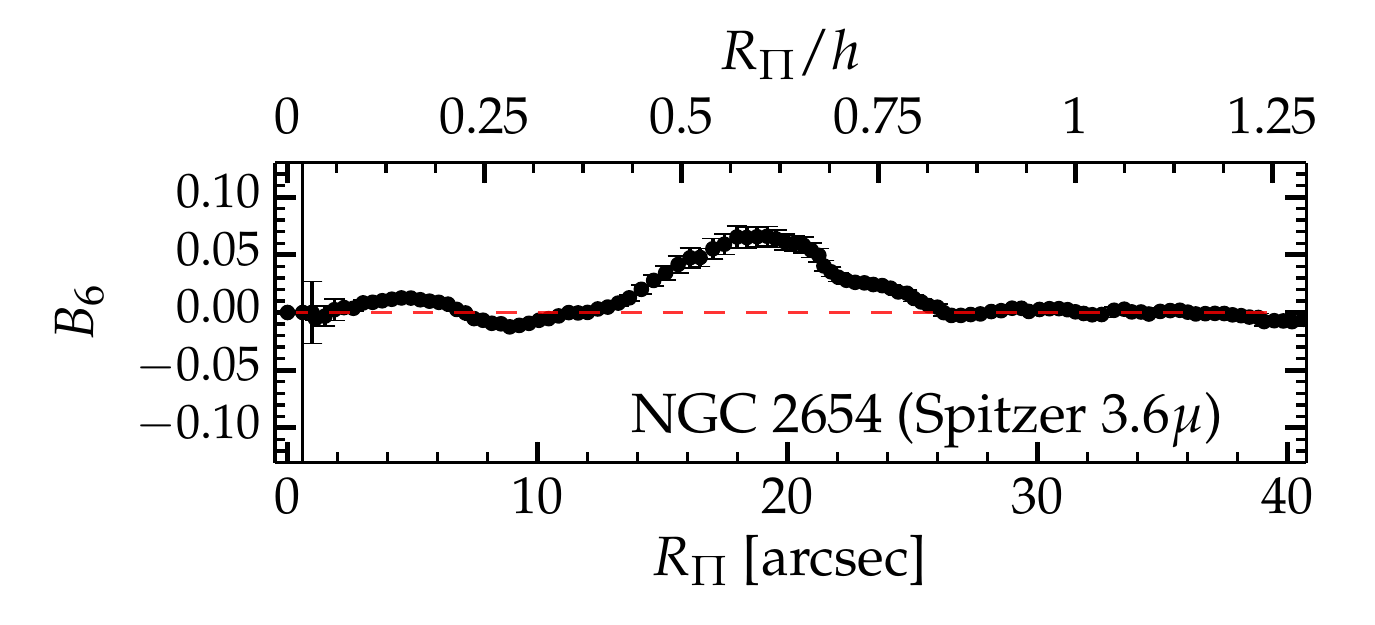} \vline & \includegraphics[width=.455\textwidth]{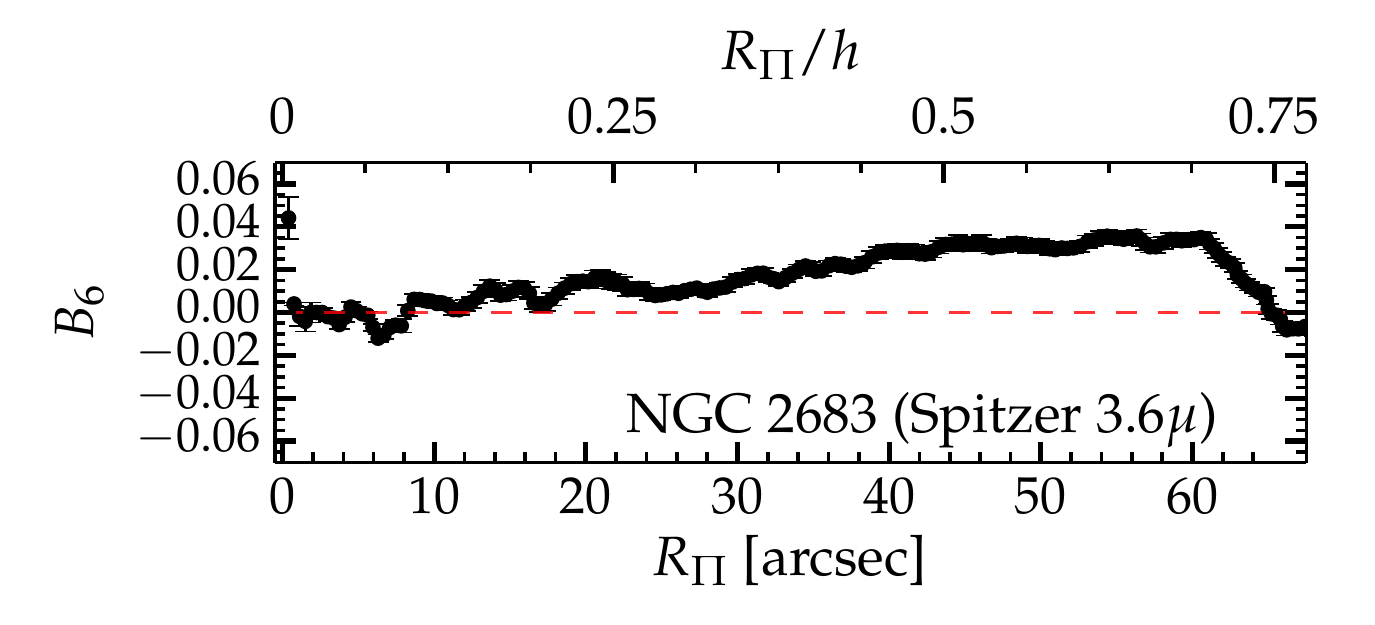}\\ 
\end{array}$
\end{center}
\label{fig:sample-1}
\end{figure*}

\setcounter{figure}{0}
\renewcommand{\thefigure}{A\arabic{figure}}

\begin{figure*}
\caption{$ - continued$.}
\centering
\begin{center}$
\begin{array}{cc}
\includegraphics[width=.455\textwidth]{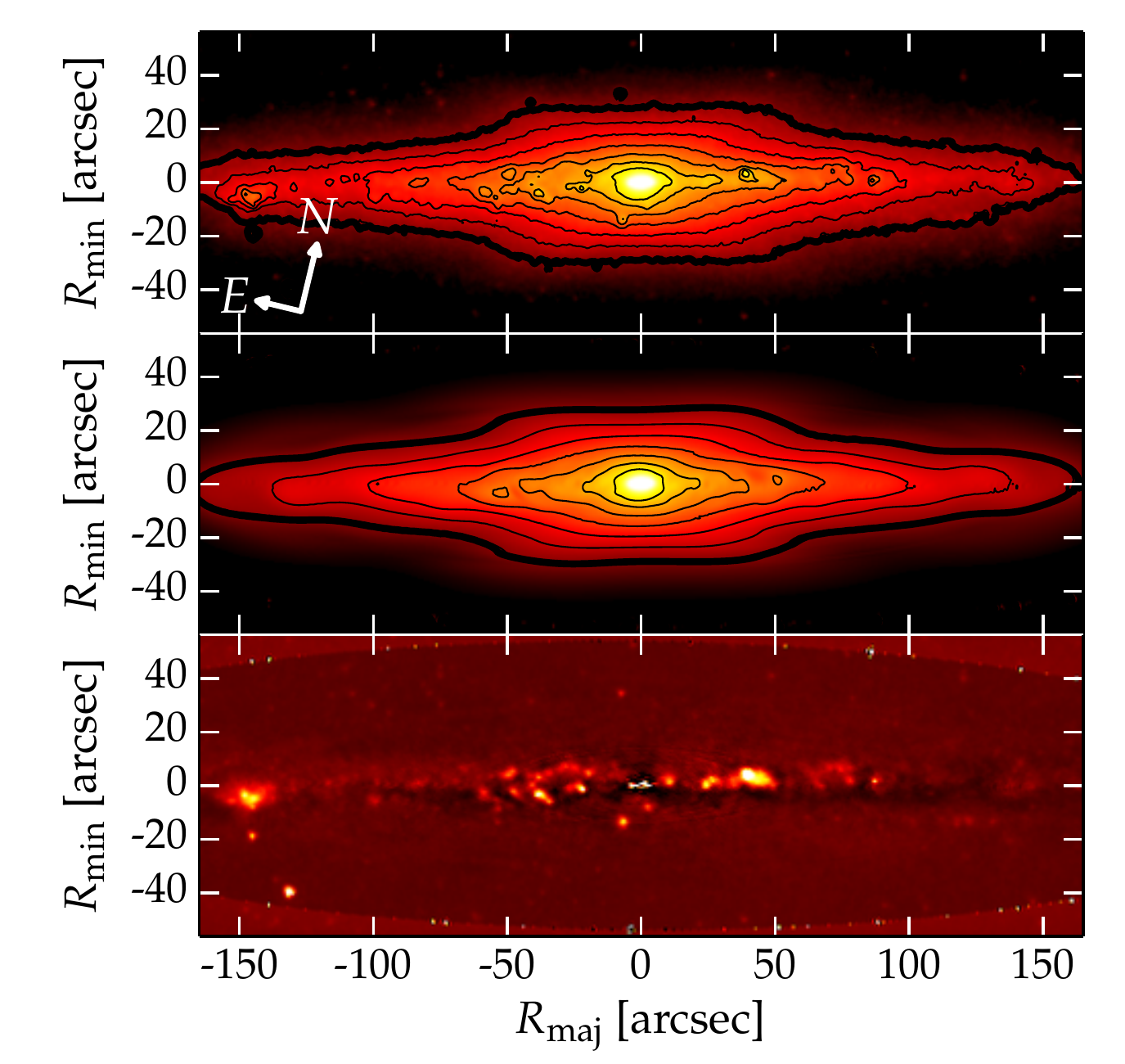} \vline & \includegraphics[width=.455\textwidth]{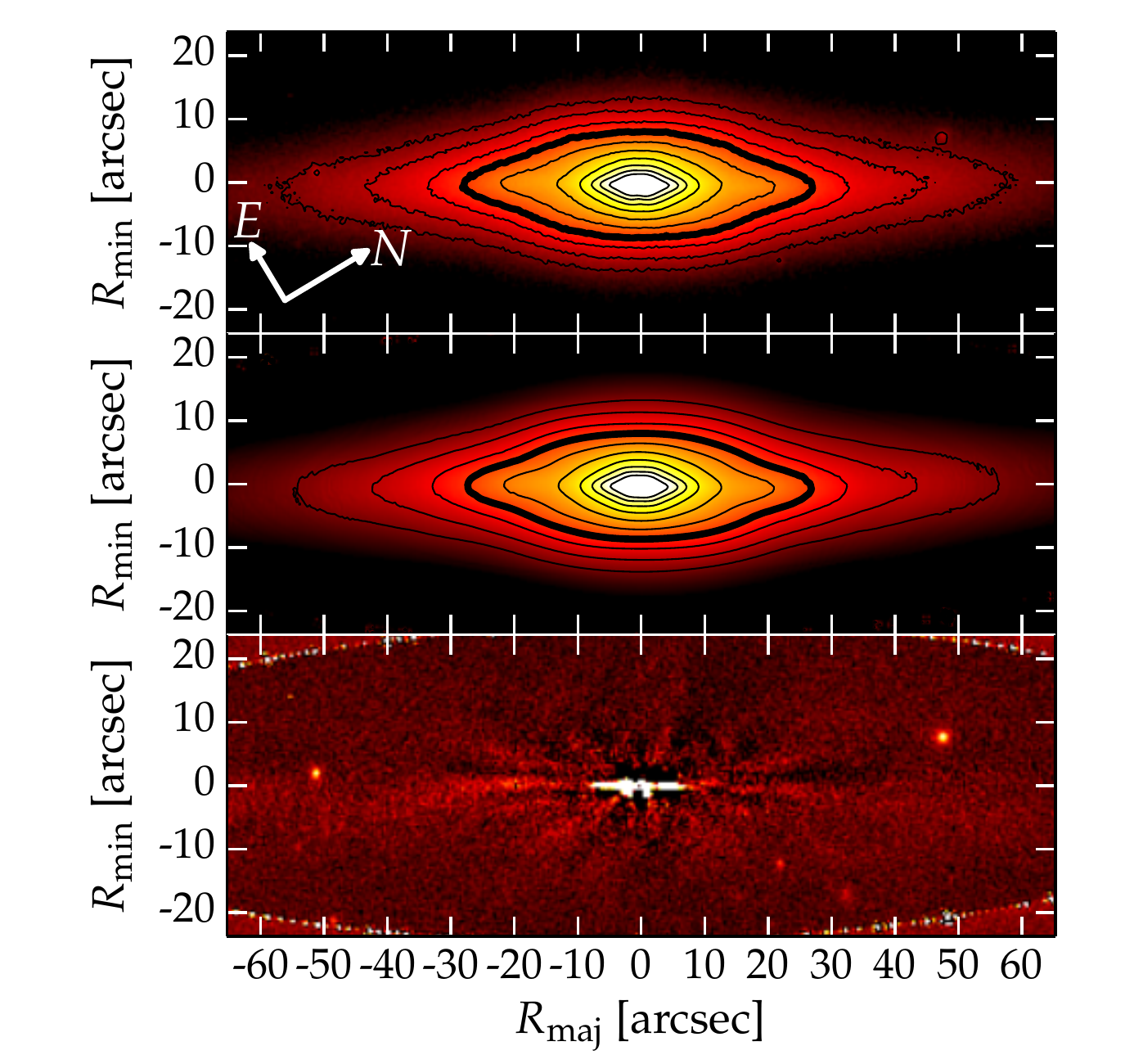} \\
\includegraphics[width=.455\textwidth]{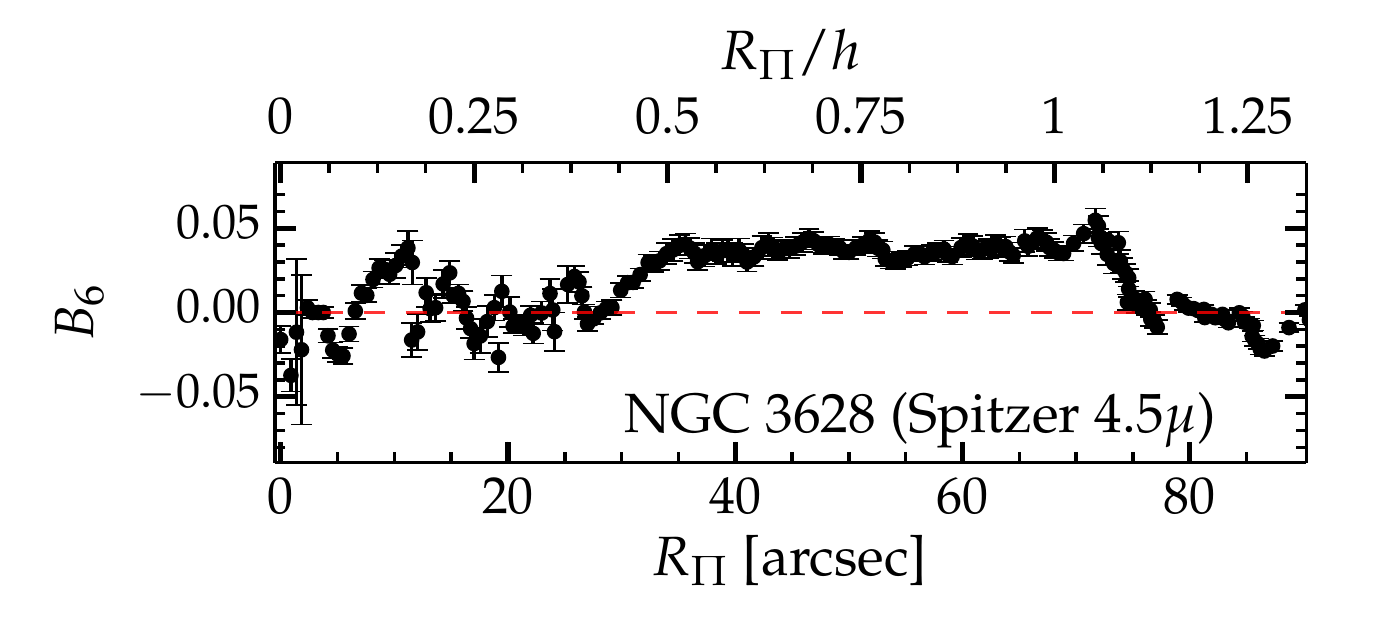} \vline & \includegraphics[width=.455\textwidth]{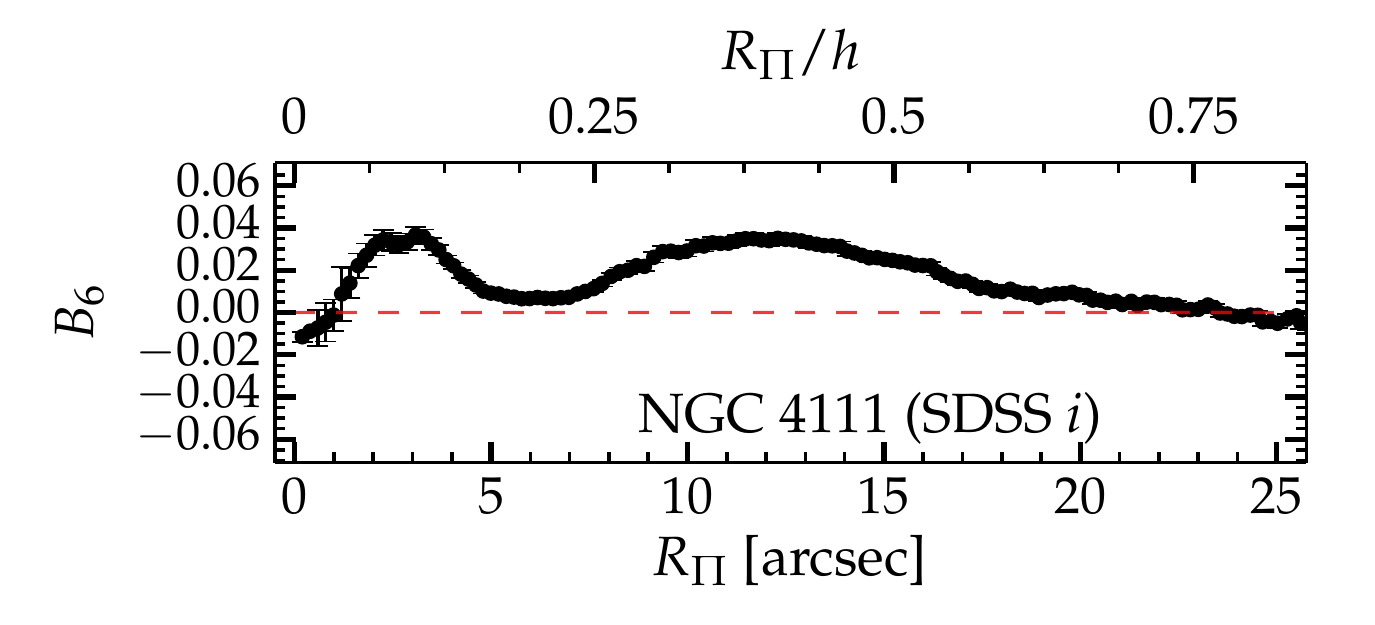}\\
\hline
\includegraphics[width=.455\textwidth]{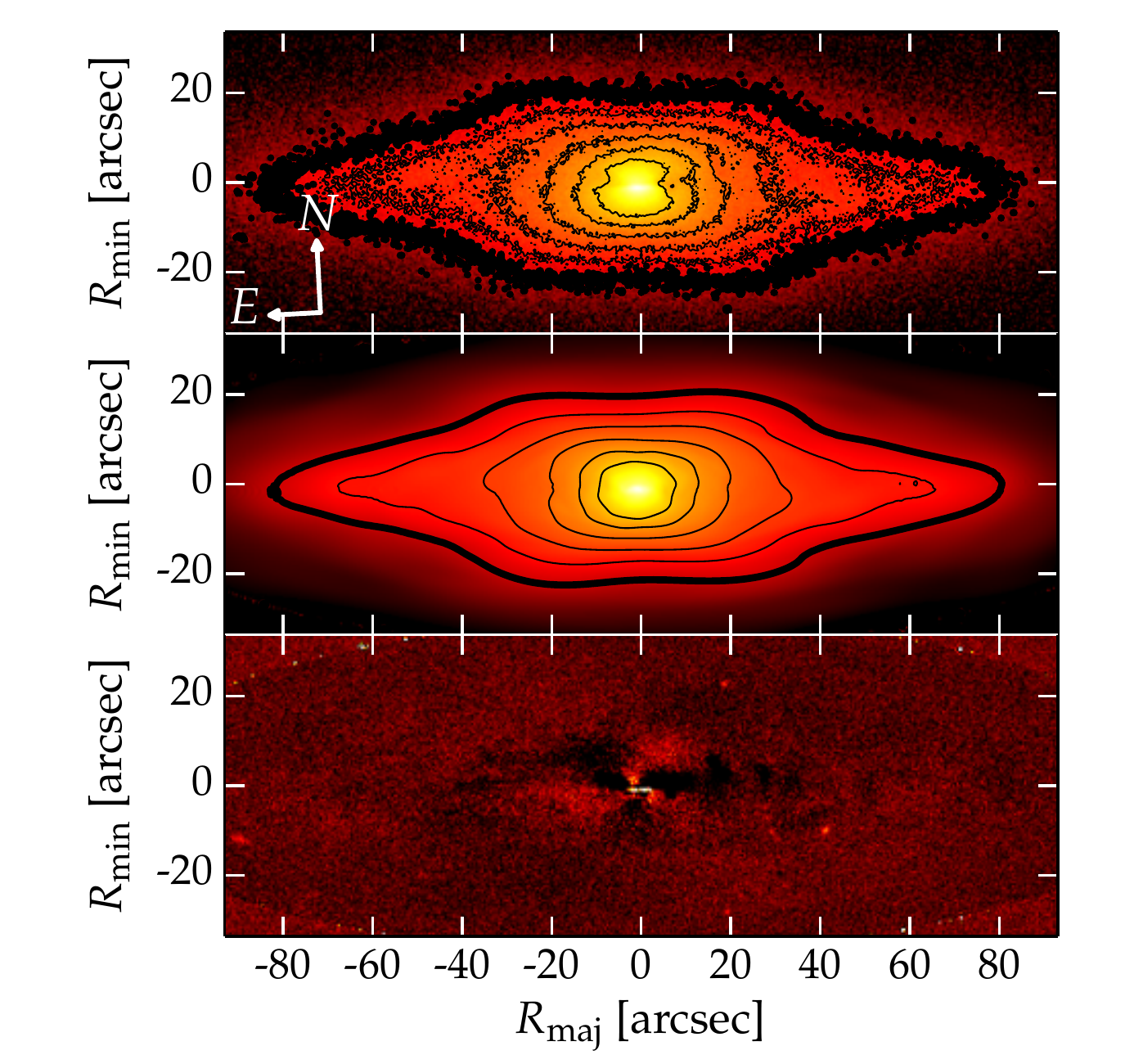} \vline & \includegraphics[width=.455\textwidth]{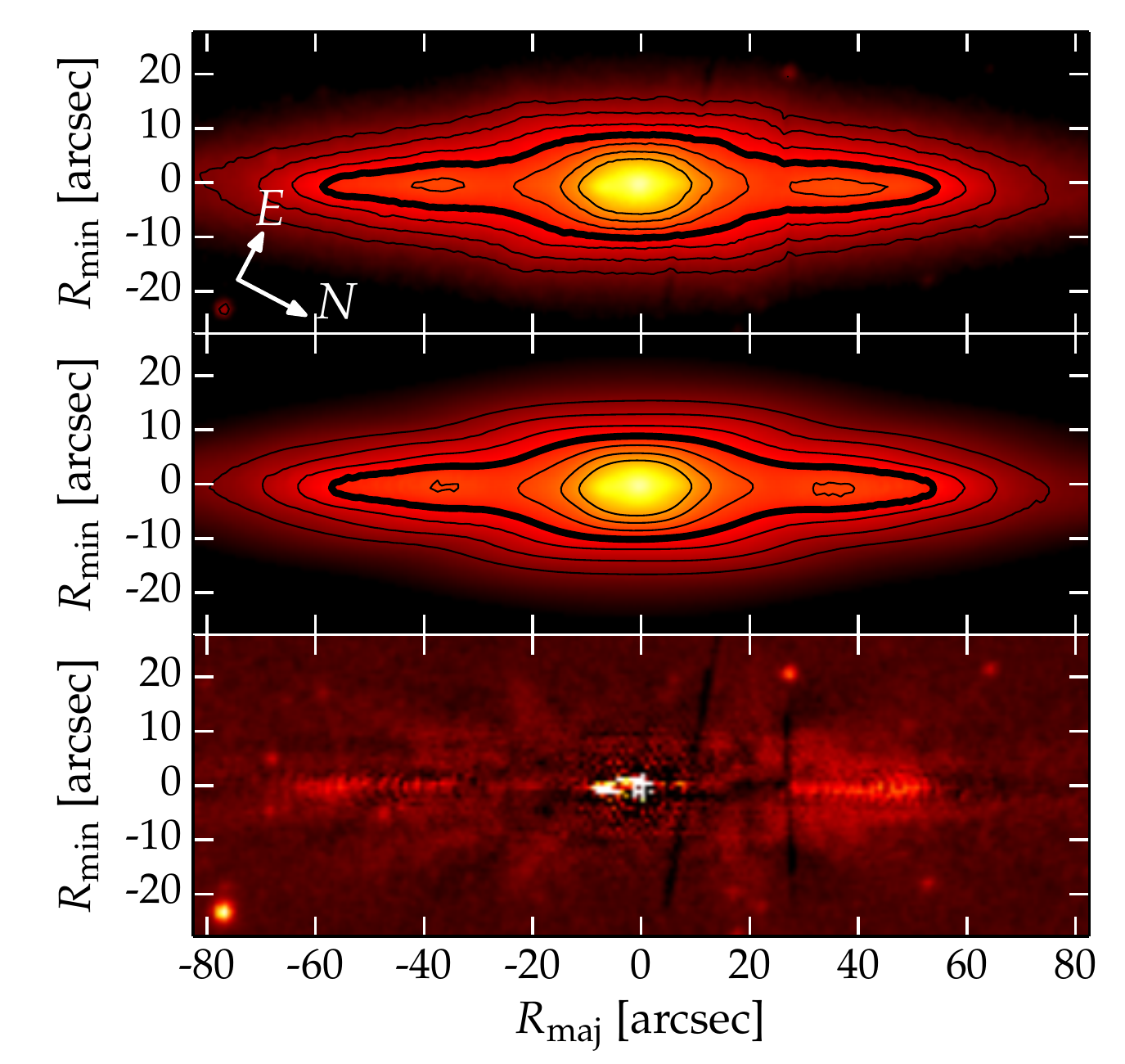} \\
\includegraphics[width=.455\textwidth]{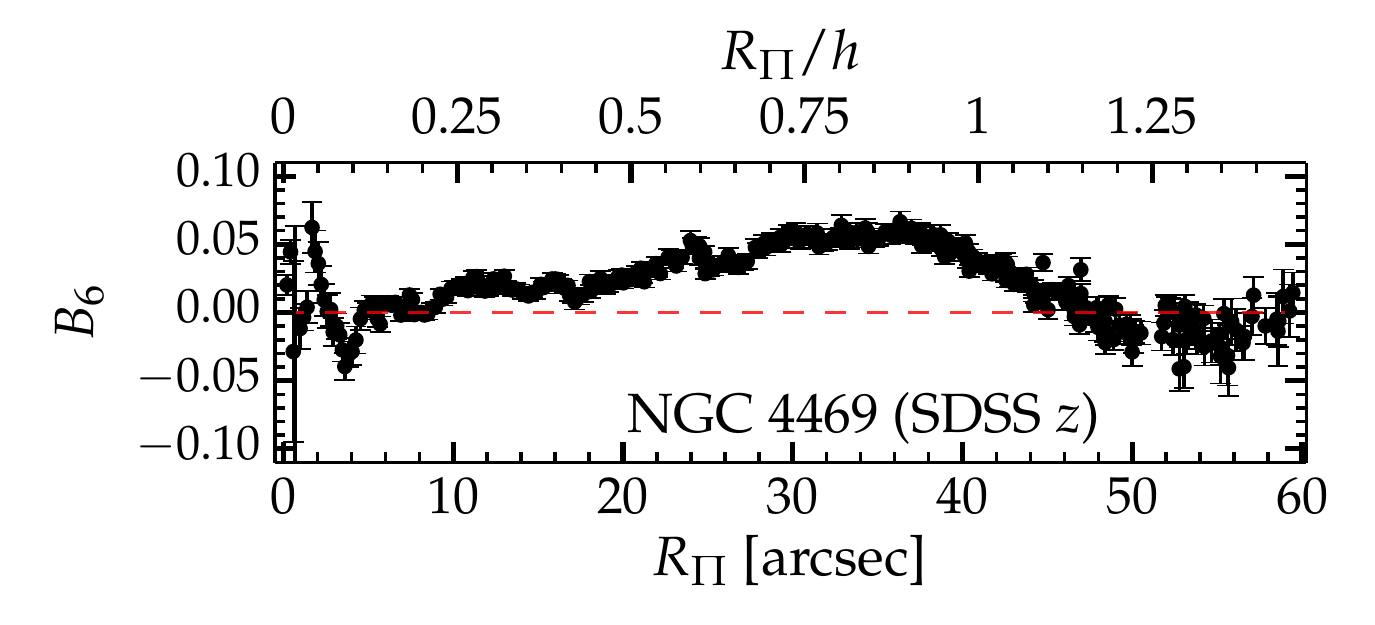} \vline & \includegraphics[width=.455\textwidth]{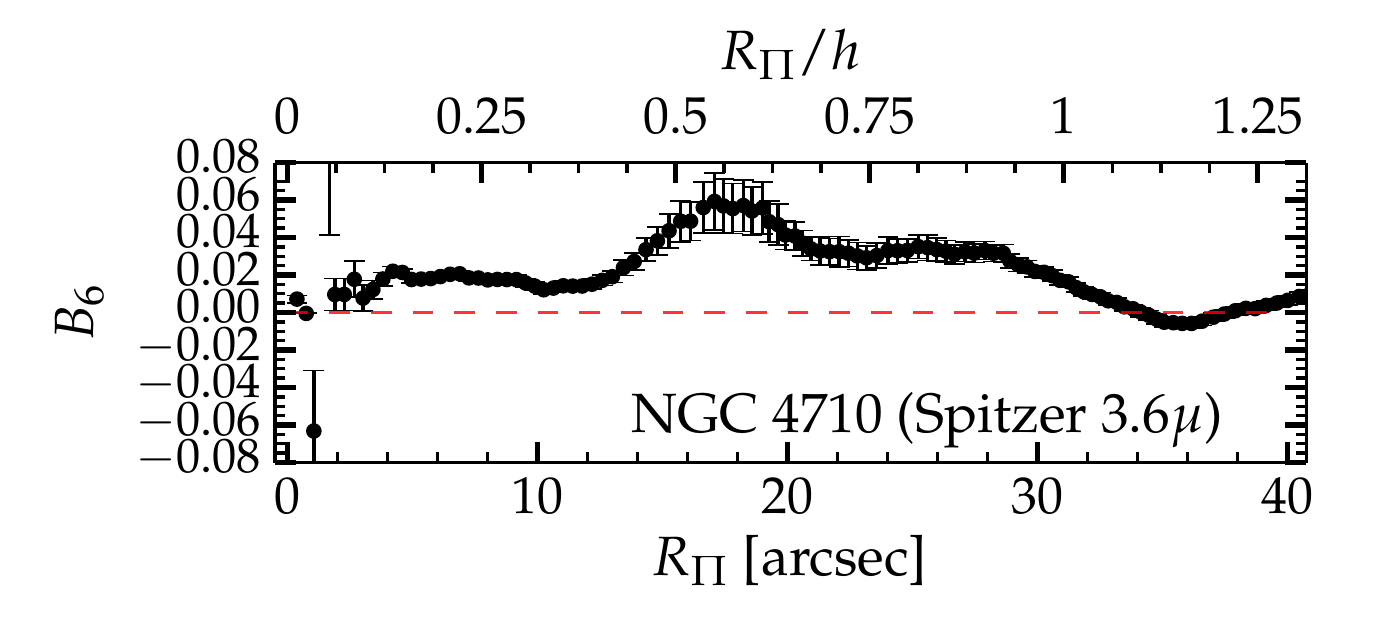}
\end{array}$
\end{center}
\label{fig:sample-2}
\end{figure*}

\setcounter{figure}{0}
\renewcommand{\thefigure}{A\arabic{figure}}

\begin{figure*}
\centering
\caption{$ - continued$.}
\begin{center}$
\begin{array}{cc}
\includegraphics[width=.455\textwidth]{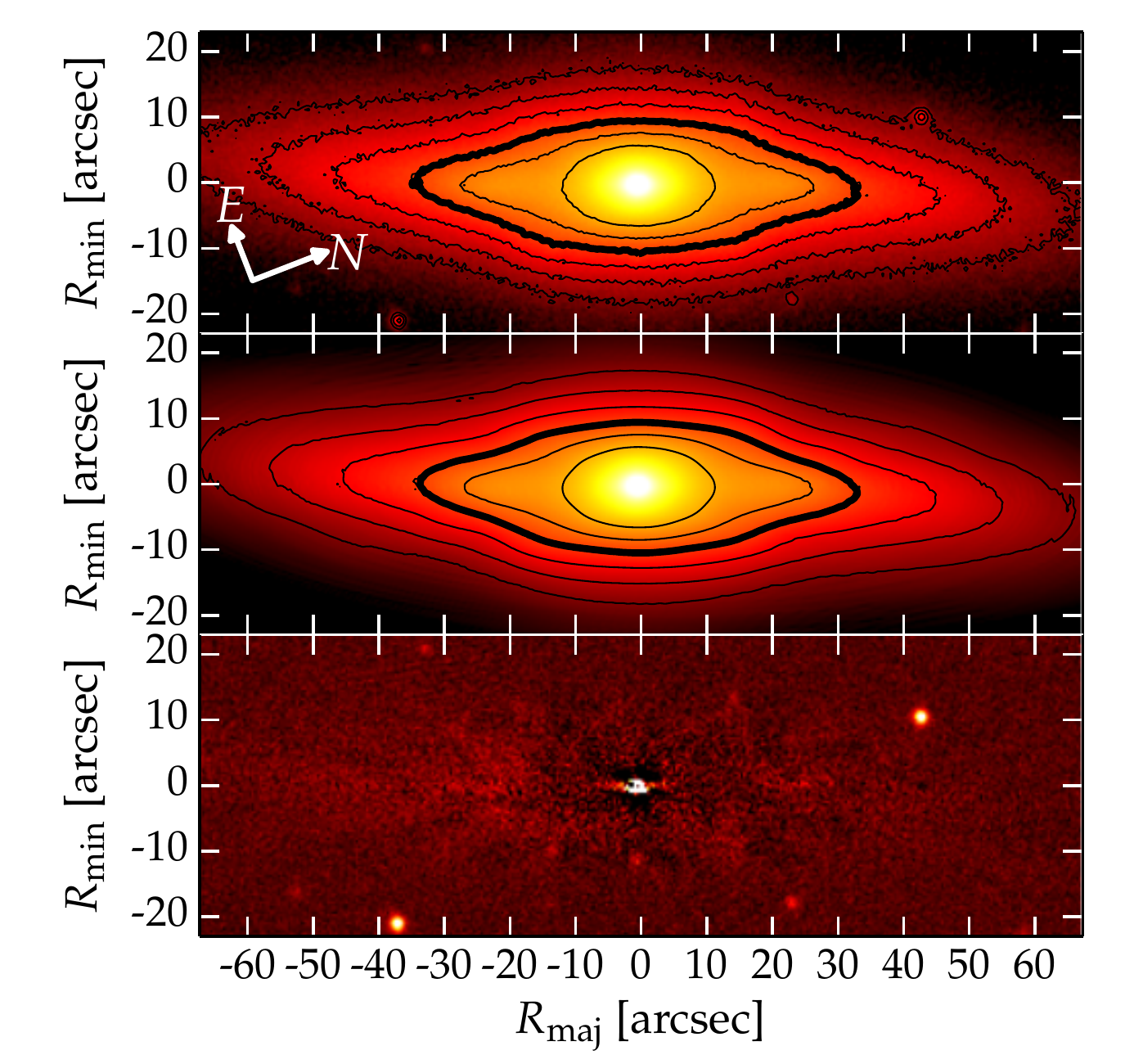} \vline & \includegraphics[width=.455\textwidth]{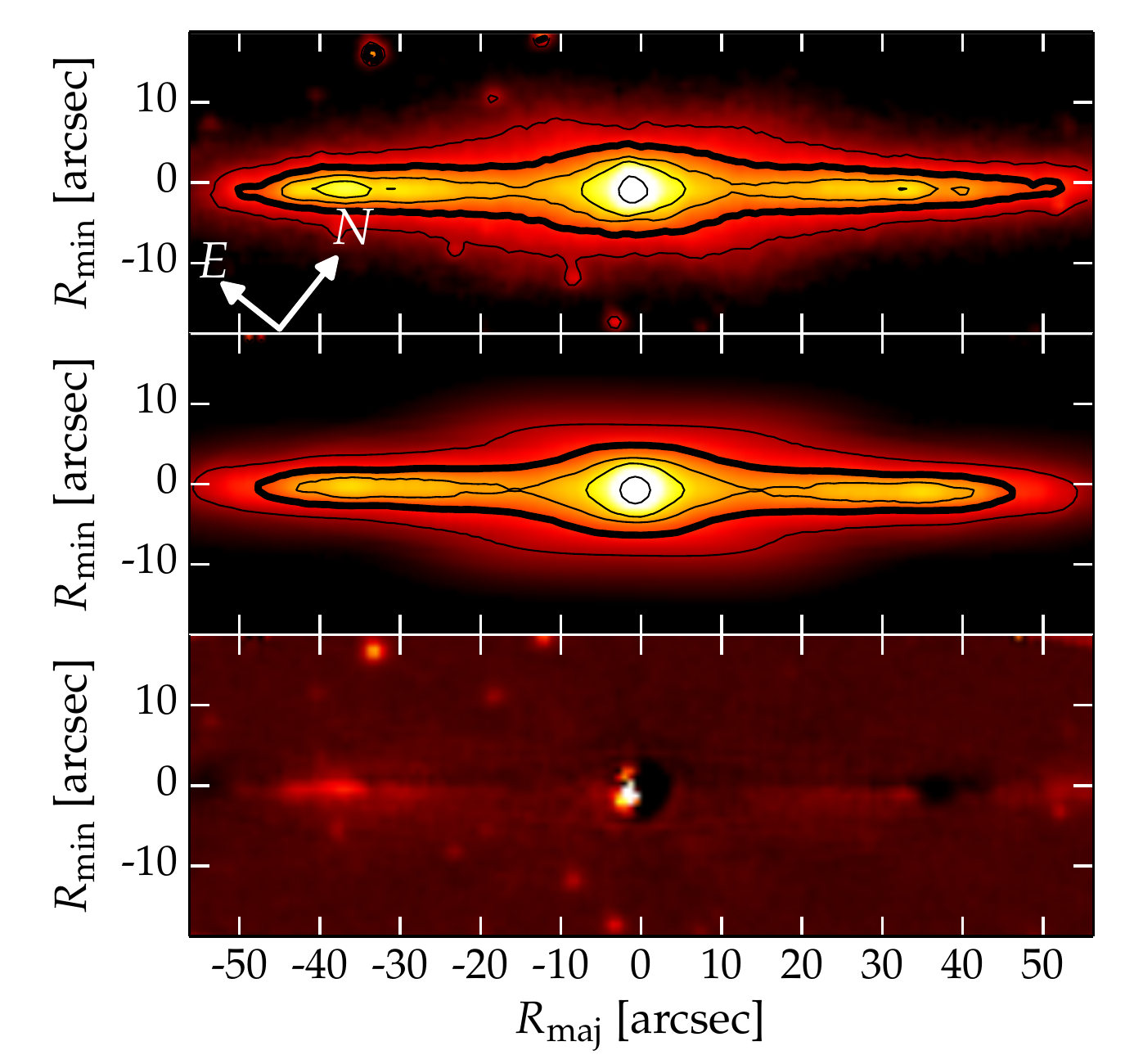} \\
\includegraphics[width=.455\textwidth]{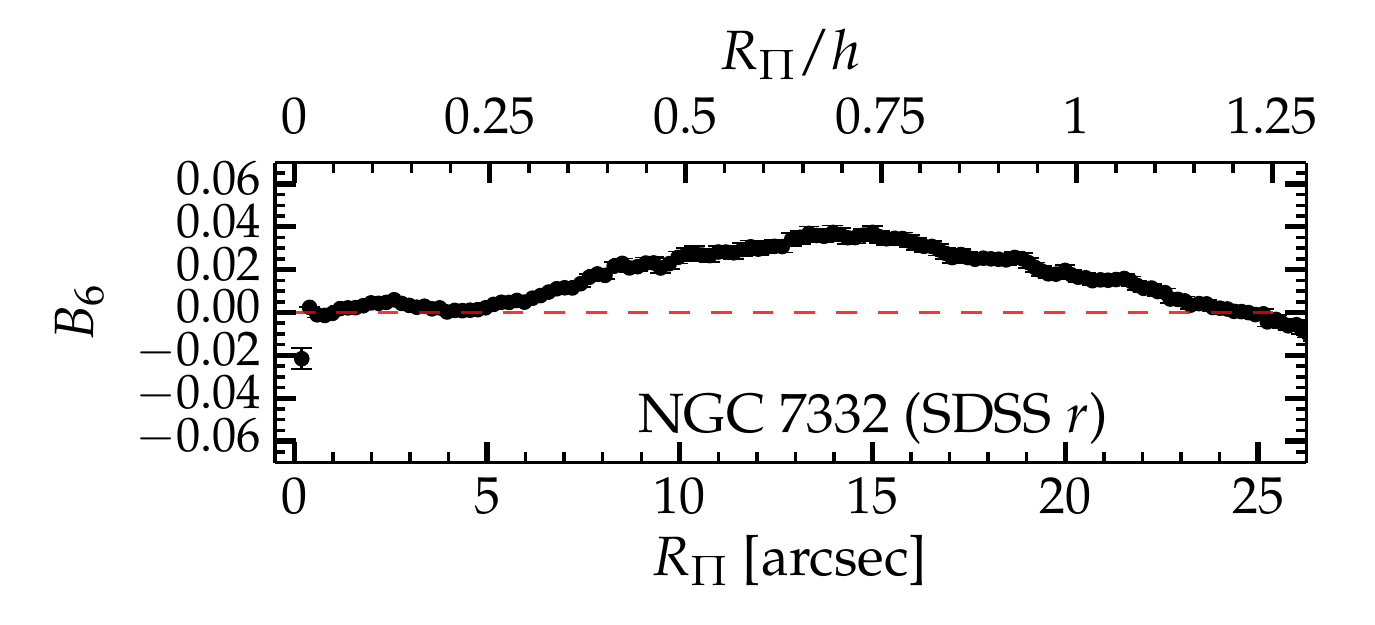}  \vline & \includegraphics[width=.455\textwidth]{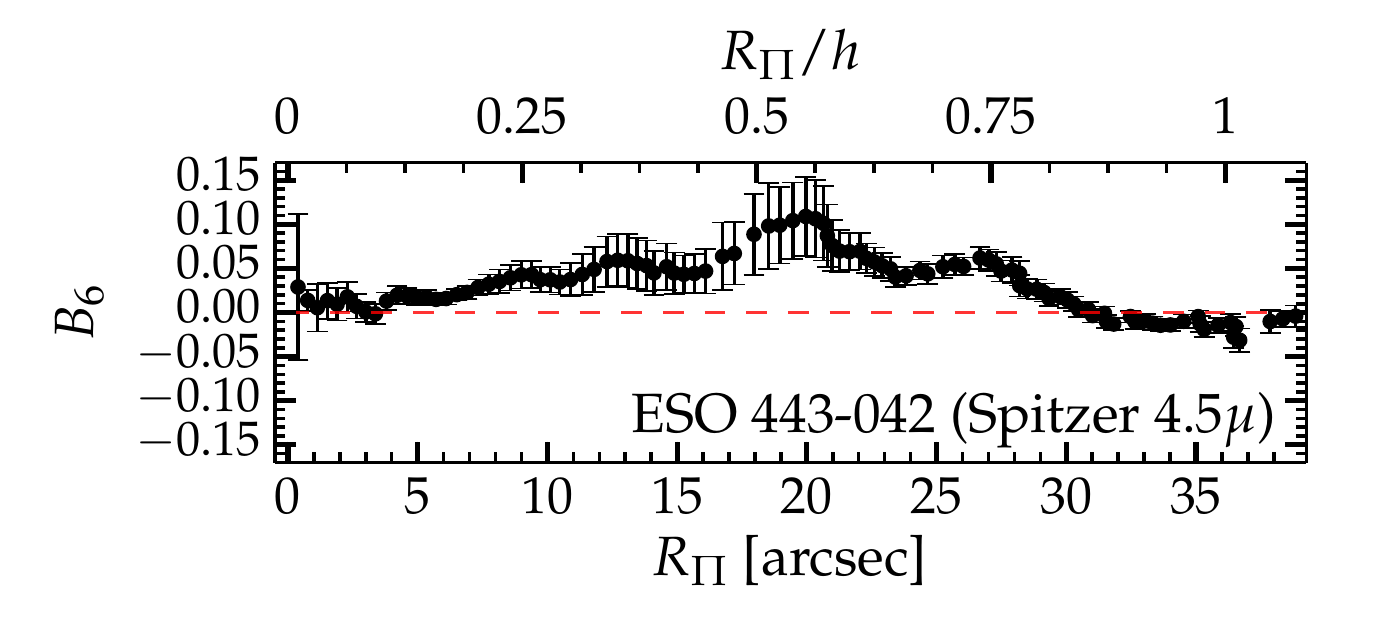}  

\end{array}$
\end{center}
\label{fig:sample-3}
\end{figure*}

\end{document}